\documentclass[prd,nofootinbib,superscriptaddress,twocolumn]{revtex4}
\usepackage[a4paper,left=1.5cm,right=1.5cm,top=3cm,bottom=3cm]{geometry}

\usepackage{amssymb,amsmath,amsfonts}
\usepackage[dvipsnames]{xcolor}
\usepackage{graphicx}
\usepackage{longtable}
\usepackage{verbatim}
\usepackage{color}
\usepackage{mdframed}
\usepackage{soul}
\usepackage{amsfonts,amssymb,mathrsfs,amsmath,esint}
\usepackage{slashed, cancel}
\usepackage{framed}
\usepackage{mdframed}
\usepackage{simplewick} 
\allowdisplaybreaks
\usepackage{balance}
\usepackage{latexsym}
\usepackage{graphicx}
\graphicspath{{./Figures/}}
\usepackage[dvipsnames]{xcolor}
\usepackage{booktabs}
\usepackage{datetime}
\usepackage{url}
\usepackage{textcomp}
\newdateformat{mydate}{\THEDAY{ }\monthname[\THEMONTH]{ }\THEYEAR}

\usepackage{tikz}
\usepackage{color}
\usepackage{framed}
\usepackage{hyperref}
\hypersetup{colorlinks, citecolor=bluscuro, linkcolor=black, urlcolor=bluscuro}
\definecolor{rossos}{cmyk}{0,1,1,0.55}
\definecolor{bluscuro}{rgb}{0.15, 0.2, .85}
\definecolor{bluchiaro}{cmyk}{1,.3,0.,0.1}

\definecolor{mygreen}{RGB}{0,130,0}

\newcommand{\beq}{\begin{equation}}
\newcommand{\eeq}{\end{equation}}
\newcommand{\bea}{\begin{eqnarray}}
\newcommand{\eea}{\end{eqnarray}}

\usepackage[normalem]{ulem}
\usepackage{xcolor}

\begin{document}

\title{\Large{Impact of correlated noise on the reconstruction of the stochastic gravitational wave background with Einstein Telescope}}

\date{\today}

 \author{Ilaria Caporali}
\affiliation{Dipartimento di Fisica ``Enrico Fermi'', Universit\`a di Pisa, Largo Bruno Pontecorvo 3, Pisa I-56127, Italy}
\affiliation{INFN, Sezione di Pisa, Largo Bruno Pontecorvo 3, Pisa I-56127, Italy}
 \author{Giulia  Capurri}
\affiliation{Dipartimento di Fisica ``Enrico Fermi'', Universit\`a di Pisa, Largo Bruno Pontecorvo 3, Pisa I-56127, Italy}
\affiliation{INFN, Sezione di Pisa, Largo Bruno Pontecorvo 3, Pisa I-56127, Italy}

\author{Walter Del Pozzo}
\affiliation{Dipartimento di Fisica ``Enrico Fermi'', Universit\`a di Pisa, Largo Bruno Pontecorvo 3, Pisa I-56127, Italy}
\affiliation{INFN, Sezione di Pisa, Largo Bruno Pontecorvo 3, Pisa I-56127, Italy}

\author{Angelo Ricciardone}
\email{angelo.ricciardone@unipi.it}
\affiliation{Dipartimento di Fisica ``Enrico Fermi'', Universit\`a di Pisa, Largo Bruno Pontecorvo 3, Pisa I-56127, Italy}
\affiliation{INFN, Sezione di Pisa, Largo Bruno Pontecorvo 3, Pisa I-56127, Italy}
\author{Lorenzo Valbusa Dall'Armi}
\affiliation{Dipartimento di Fisica ``Enrico Fermi'', Universit\`a di Pisa, Largo Bruno Pontecorvo 3, Pisa I-56127, Italy}

\begin{abstract}
\noindent
Einstein Telescope (ET) is a proposed next-generation Gravitational Wave (GW) interferometer designed to detect a large number of astrophysical and cosmological sources with unprecedented sensitivity.  A key target for ET is the detection of a stochastic gravitational-wave background (SGWB), a faint signal from unresolved GW sources. In its proposed triangular configuration, correlated Newtonian noise of seismic origin poses some challenges for the SGWB detection. We study the impact of correlated noise on the SGWB detection and relative parameter estimation for ET in the triangular configuration, comparing it to a 2L configuration with two separated L-shaped detectors. We perform a Bayesian analysis on simulated data, which shows that accurate reconstruction of the SGWB parameters and instrumental noise is achievable if the noise is properly modeled. 
We illustrate that neglecting correlated noise leads to significant biases in the parameter reconstruction. Our results show that while the 2L configuration provides slightly better parameter estimation precision, mainly due to its longer arm length, the triangular configuration remains competitive when accurate noise modeling is provided.

\end{abstract}

\maketitle

\section{Introduction} 

Einstein Telescope (ET) represents the next generation of Gravitational Wave (GW) observatories, promising unprecedented sensitivity and the capability to detect a vast range of astrophysical and cosmological sources~\cite{Punturo:2010zz, ET:2019dnz}. Among its primary scientific objectives, ET aims to detect the stochastic gravitational wave background (SGWB), a faint and diffuse signal originating from the superposition of unresolved GW sources, both of astrophysical and cosmological origin. The detection of the SGWB would provide invaluable insight into the population of compact binaries throughout the universe, as well as potential signals from the early universe~\cite{Regimbau:2011rp,Caprini:2018mtu}. \\
One of the main challenges in the search for the SGWB is the detailed understanding of the noise properties of the detector, in order to distinguish it from the signal. For this reason, many preliminary studies are conducted to characterize how site-dependent properties of the noise can impact on the science objectives~\cite{Janssens:2022tdj, Janssens:2023anf, Janssens:2024jln, di2021seismological, andric2020simulations}. 
According to the Conceptual Design Report~\cite{ETdesignRep}, ET should be an underground giant nested triangular detector of 10 km arms, with three detectors dedicated to the low frequency regime and three to the high frequency one, in a xylophone configuration. Other possible configurations, as two separated L-shape detectors of different arm-lenghts, have been recently compared to the triangular configuration~\cite{Branchesi:2023mws}. Being a new layout, with relevant advancements in technology, when the triangular configuration is considered, fair and robust comparisons require the inclusion of all the source of noise and geometry properties. Recently, in~\cite{Janssens:2022xmo}, it has been noticed that in the triangular configuration, some noise correlations among the detectors appear. In fact, since some mirrors of different nested interferometers are apart by a few hundred meters, on this scale appear correlated seismic and Newtonian noise~\cite{Janssens:2024jln}. According to such analysis, correlated noise would impact the sensitivity to the SGWB of ET, in its 10 km triangle configuration, up to frequencies of order (50 - 100) Hz. This, of course, would have large implications on the detection of the SGWB. The impact of such a noise correlation on the parameter estimation of resolved events has recently been studied in~\cite{Cireddu:2023ssf}, while the impact of magnetic noise correlation for separated detectors has been studied in~\cite{Janssens:2022tdj}. However, to the best of our knowledge, a careful analysis for the SGWB in the case of the triangular configuration, has not been addressed so far. \\
In this paper, we study how correlated noise impacts the detection and characterization of the SGWB in the ET triangular configuration, and we compare it to the 2L configuration. SGWB searches with current ground-based detectors consist in cross-correlating the output of multiple detectors~\cite{Romano:2016dpx}.  When we have a triangular detector, we can make use of the data from three detectors (or channels). A common choice is to work on the XYZ basis or in the AET where the noise matrix is diagonal and a channel can be built where the GW signal is strongly suppressed compared to instrumental noise (T or null channel)~\cite{Hogan:2001jn, Adams:2010vc,Janssens:2022cty}. 
We work in the latter basis and we perform a Bayesian analysis for a power-law SGWB in the presence of correlated noise. We derive the likelihood for the SGWB estimator both in the case of correlated detectors (i.e., triangular) and for the uncorrelated ones (i.e., 2L). In this way, we study the reconstruction of the SGWB signal in presence of correlated noise. We show that, even in the presence of correlated noise, we are able to accurately reconstruct the parameters of the SGWB. Moreover, we do not limit our analysis to the reconstruction of the SGWB parameters, but we also estimate the parameters describing the instrumental noise. We compare the accuracy on the reconstructed parameters, both signal and noise, for the two configurations. Finally, we study the impact of neglecting the correlated noise in the SGWB reconstruction by analyzing simulated data that include correlated noise using a likelihood that does not account for it. We highlight the presence of a bias that would be introduced into the inferred SGWB parameters if noise correlations are not properly accounted for, providing crucial insights into the importance of accurate noise modeling. Furthermore, we validate the robustness of our Bayesian analysis using a probability-probability (PP) plot, which helps to confirm the reliability of our findings. \\

The structure of the paper is organized as follows. Section \ref{sec:Section2} introduces the formalism used to describe a SGWB. Section \ref{section:ET configurations and noise model} focuses on characterizing the noise in two configurations: the triangular configuration (detailed in Subsection \ref{sec:Characterization of the noise for the triangular configuration}) and the 2L configuration (discussed in Subsection \ref{sec: Characterization of the noise for the 2L configuration}). Particular emphasis is placed on the presence of correlated Newtonian noise in the triangular configuration. Section \ref{sec:bayesian analyis} explains the Bayesian analysis framework. Specifically, Subsection \ref{subsec:likelihood} outlines the likelihood function employed, Subsection \ref{subsec:data_generation} explains the data generation procedure implemented for the analysis, and, finally, Section \ref{sec:Parameter estimation outcomes} presents the parameter estimation results.

\section{Stochastic Gravitational Wave Background}
\label{sec:Section2}
A SGWB can be characterized as a superposition of GWs coming from all directions \cite{Allen:1997ad,Romano:2016dpx},
\begin{equation}
    h_{ij}(t,\textbf{x})=\int df\, d\hat{n} \sum_{\lambda} \tilde{h}_\lambda(f,\hat{n})e_{ij}^\lambda(\hat{n})e^{2\pi i f(t-\hat{n}\cdot \textbf{x}/c)} \, ,
\end{equation}
where $\hat{n}$ is the propagation direction, $\lambda= +, \times$ labels the two polarizations, $e_{ij}^\lambda$ are polarization tensors and $\tilde{h}_{\lambda}$ the wave amplitude of the background. 
For a SGWB that is stationary, isotropic and unpolarized, the two-point correlation function can be written as
\begin{equation}
    \langle \tilde{h}_\lambda(f, \hat{n}) \tilde{h}_{\lambda'}^*(f', \hat{n}')\rangle = \frac{\delta(f-f')}{2} \frac{\delta_{\hat{n} \hat{n}'}}{4\pi} \delta_{\lambda \lambda'} \frac{3 H_0^2}{4 \pi^2 f^3} \Omega_{\rm GW}(f) \, ,
    \label{eq:twopoint_corr}
\end{equation}
where the $\delta_{\lambda \lambda'}$ enforces the unpolarized nature, $\delta_{\hat{n} \hat{n}'}$ the isotropy, and $\delta(f-f')$ the stationarity of the background. $H_0$ represent the Hubble parameter today and $\Omega_{\rm GW}$ represents the GW energy density per logarithmic interval of frequency.\\
The spectral dependence of the SGWB energy density is a key observable for distinguishing among several sources~\cite{ Caprini:2019pxz, Flauger:2020qyi,LISACosmologyWorkingGroup:2022jok,Braglia:2024kpo, Blanco-Pillado:2024aca,Caprini:2024hue}. For the astrophysical background generated by the superposition of many compact binary coalescences (CBCs),  the spectral tilt is predicted to be $n_{\rm GW} = 2/3$~\cite{Phinney:2001di} (at least for the inspiral phase). Moreover, this background is expected to be Gaussian~\cite{Regimbau:2011rp} and unpolarized, up to small fluctuations in the circular and linear polarization of the order of $10^{-2}-10^{-3}\, \Omega_{\rm GW}$~\cite{ValbusaDallArmi:2023ydl,Belgacem:2024ohp}. The  cosmological background predicted by the simplest single-field inflationary models is expected to be nearly flat ($n_{\rm GW} \simeq 0$) with a slight red tilt~\cite{Grishchuk:1974ny, Starobinsky:1979ty,Guth:1980zm,Starobinsky:1980te,Linde:1981mu,Albrecht:1982wi}, with an amplitude which lies below the sensitivity of any current or future generation interferometer~\cite{Caprini:2018mtu, Guzzetti:2016mkm}. However, larger amplitudes and more complex spectral shapes—such as broken power laws or Gaussian bumps—may arise in multi-field inflationary scenarios~\cite{Sorbo:2011rz, Badger:2024ekb, Garcia-Bellido:2023ser, Biagetti:2013kwa, Cook:2011hg, Barnaby:2012xt},
effective field theories of inflation \cite{Bartolo:2015qvr,Ricciardone:2016lym,   Capurri:2020qgz}, scalar-induced gravitational waves~\cite{Tomita:1975kj, Matarrese:1992rp,Matarrese:1993zf,Matarrese:1997ay, Acquaviva:2002ud, Mollerach:2003nq,Ananda:2006af, Baumann:2007zm, Domenech:2021ztg,Perna:2024ehx,Iovino:2024sgs}, first-order phase transitions~\cite{Witten:1984rs, Hogan:1986qda, Caprini:2009fx, Kamionkowski:1993fg, Caprini:2007xq, Huber:2008hg, Caprini:2009yp, Hindmarsh:2013xza}, or cosmic strings~\cite{Caprini:2024ofd, Nielsen:1973cs, Sakellariadou:2009ev, Vilenkin:1981bx, Sakellariadou:1990ne, Hindmarsh:1990xi, Damour:2004kw}.

In our analysis, we focus on a power-law behaviour and parametrize the SGWB energy density as
\begin{equation}
    \Omega_{\rm GW}(f) = A_{\rm GW}\, \left(\frac{f}{25\, \rm Hz}\right)^{n_{\rm GW}} \, ,
    \label{eq:sgwb_parametrization}
\end{equation}
where $A_{\rm GW}$ is the amplitude at the reference frequency $25\, \text{Hz}$, and $n_{\rm GW}$ is the spectral tilt. 
The Fourier transform of the data stream measured by an interferometer, denoted as $\tilde{s}_I(f)=\tilde{h}_I(f)+\tilde{n}_I(f)$, consists of the sum of the GW stochastic signal and the noise. The data is assumed to be `perfect' residuals, meaning that all transients including loud deterministic signals and glitches in the noise are assumed to be subtracted from the time stream through some appropriate methods.
The projected GW amplitude on the detector $I$, $\tilde{h}_I(f)$, is related to the GW amplitude $\tilde{h}_\lambda$ through the detector's pattern function $F_I^\lambda$
\begin{equation}
    \tilde{h}_I(f) = \int d\hat{n} \sum_\lambda F_I^{\lambda}(f,\hat{n})\, \tilde{h}_\lambda(f,\hat{n}) \, .
    \label{eq:GWsignal_Fourier}
\end{equation}
The SGWB signal have a zero mean and  covariance given by~\cite{Allen:1997ad,Romano:2016dpx}
\begin{equation} 
\left\langle \tilde{h}_I(f)\tilde{h}^*_J(f^\prime)\right\rangle \equiv \frac{\delta(f-f^\prime)}{2}\frac{3H_0^2}{10\pi^2f^3} \gamma_{IJ}(f) \Omega_{\rm GW}(f) \, , \label{eq:2pt} 
\end{equation}
where the quantity $\gamma_{IJ}(f)$ represents the normalized overlap reduction function (ORF), defined as \cite{Thrane:2013oya, Romano:2016dpx}
\begin{equation}
\gamma_{IJ}(f) \equiv \frac{5}{8\pi} \int d\hat{n} \sum_\lambda F_I^\lambda(f, \hat{n}) F_J^{\lambda, *}(f, \hat{n})  e^{-2\pi i f \hat{n} \cdot (\textbf{x}_I - \textbf{x}_J)/c} \, . 
\label{eq:orf_general_basis}
\end{equation} 
The indices $I,J$ run over the interferometer channels in the case of the triangular configuration, whereas for the 2L configuration, they correspond to the two detectors.
The ORF is normalized to unity for two co-located, L-shaped detectors.

\section{ET configurations and noise model}
\label{section:ET configurations and noise model}

ET is a proposed third generation interferometer, which, according to the technical design report~\cite{ETdesignRep}, will have a triangular layout with three nested detectors in a xylophone configuration, located 200-300 meters underground. Each detector will consist of two interferometers: one optimized for the high-frequency and the other for the low-frequency. In the triangular configuration, each interferometer arm will be 10 km long. Recently, an alternative design for the ET has been considered, involving two separate L-shaped detectors located at different sites, with each interferometer arm ranging from 10 to 20 km in length and oriented differently to maximize coverage. A quantitative analysis comparing these configurations has been conducted in~\cite{Branchesi:2023mws}.

Potential locations for the ET are the area near the Sos Enattos mine in Sardinia, Italy, and the Meuse-Rhine three-border region across Belgium, Germany and the Netherlands. Recently, also Saxony has been proposed as an alternative site~\cite{saxony_einstein_telescope_2024}.

In this paper, we consider ET in the triangular configuration located in Sardinia and ET in the 2L configuration with one detector located in Sardinia and one in the Meuse-Rhine region. We consider the two separated detectors in the {\it{aligned}} configuration, so using the relative orientation between the two, defined with respect to the great circle that connects them. \footnote{We use the 2L configuration with aligned detectors, as it maximizes the sensitivity to the SGWB. 
However, another possibility is a misaligned configuration with a mutual orientation of 47.51° (see \cite{Branchesi:2023mws} for the impact on the SGWB).}

The impact of different ET configurations on the scientific objectives is an area of active research (see e.g.,~\cite{Branchesi:2023mws,Iacovelli:2024mjy, Ebersold:2024hgp}).  Beyond the geometry, reducing and characterizing the various noise sources affecting ET's sensitivity is crucial for achieving the expected scientific outcomes. One of the main goals for ET is to improve sensitivity at low frequencies. In this regard, considerable effort in recent years has focused on understanding the effects of seismic and Newtonian noise on ET's sensitivity \cite{Koley:2022wpe, andric2020simulations, di2021seismological}.  This challenge is heightened for the triangular configuration, which is potentially threatened by correlated seismic and Newtonian noise~\cite{Saulson:1984yg, Badaracco:2019vjq}.  This correlation arises because the input and output mirrors of two interferometers are proposed to be separated by only a few hundred meters. As noted in \cite{Janssens:2022xmo}, there are five specific coupling points between two of the three interferometers. Although direct seismic noise would affect the search for an isotropic SGWB only up to around 5 Hz~\cite{Janssens:2023anf}, the challenge lies in its ability to induce variations in the local gravitational field, a phenomenon known as Newtonian noise (NN). Two types of seismic waves contribute to the NN~\cite{Janssens:2023anf, Koley:2022wpe}: Rayleigh waves, which propagate along the Earth's surface, and body waves, which travel through the Earth's mantle. Rayleigh waves diminish rapidly with depth, while body waves are expected to be a more substantial source of correlated noise for colocated detectors, since they will be located several hundred meters underground. This type of noise could affect the isotropic SGWB searches at frequencies up to approximately 40 Hz~\cite{Janssens:2022xmo, Janssens:2024jln, Branchesi:2023mws}. For these reasons, correlated noise sources could limit the detectability of an isotropic SGWB for ET in its triangular configuration.\footnote{There is also correlated noise from magnetic field fluctuations \cite{Janssens:2021cta,Janssens:2022tdj}, which is the only known broadband environmental noise source that could affect far-away detectors (i.e., 2L) because it coherently couples to widely separated, earth-based interferometers. This noise arises from Schumann resonances—electromagnetic waves along Earth’s surface generated by global lightning activity. However, in this paper, we disregard the impact of this correlated noise source and we leave the analysis for future works.}

\subsection{Characterization of the noise for the triangular configuration}
\label{sec:Characterization of the noise for the triangular configuration}

In the case of ET in a triangular configuration, we will measure data $s_I(t)$ in each of the three different channels. In general, the output of a given channel, labeled as $I$, can be written as
\begin{equation}
    s_I(t) = h_I(t) + n_I(t) \, ,
    \label{eq:data_time_domain}
\end{equation}
where $h_I(t)$ and $n_{I}(t)$ are the time-domain signal and noise contributions, respectively.
Under the assumption that every transient signal has been removed, the data have zero mean and covariance given by 
\begin{equation}
    \Sigma_{IJ}(t,t^\prime) \equiv \left\langle s_I(t)s_J(t^\prime)\right\rangle \, .
\end{equation}
If we consider only stationary noise and SGWB, the covariance matrix depends only on $|t - t^\prime|$. By definition, the covariance matrix is also symmetric under the exchange of two channels, $\Sigma_{IJ}=\Sigma_{JI}$. Therefore, for a fixed pair of detector channels, the covariance is a symmetric Toeplitz matrix, i.e., a matrix with constant values on each ascending diagonal. In~\cite{zhu2017asymptotic}, it has been shown that a Toeplitz matrix is asymptotically equivalent to a circulant matrix, whose eigenvalues are the Discrete Fourier Transform (DFT) of the covariance (see also~\cite{Cireddu:2023ssf} for a more detailed discussion). This property implies that the DFT basis allows us to diagonalize the covariance matrix in the time domain by using the DFT of the data as a diagonal basis. For this reason, from now on, we will focus exclusively on the Fourier transform of the data, assuming that the DFT of the time delays has already been performed. 
In particular, we denote the Fourier transform of the SGWB signal, defined in Eq.~\eqref{eq:GWsignal_Fourier}, as $\tilde{h}_I(f)$, and the Fourier transform of the noise as $\tilde{n}_I(f)$.

Considering two detectors $I$ and $J$,
the covariance of the noise is defined as the two-point correlation function
\begin{equation}
    \left\langle \tilde{n}_I(f)\tilde{n}^*_J(f^\prime)\right\rangle \equiv \frac{\delta(f-f^\prime)}{2} N_{IJ}(f) \, ,  
    \label{def:PSD_noise}
\end{equation}
where we have assumed that the noise is stationary and, therefore, not correlated at different frequencies. We also assume that the noise is Gaussian~\cite{Caprini:2019pxz, Flauger:2020qyi}.
In Eq.~\eqref{def:PSD_noise}, the elements along the diagonal, i.e., those with $I=J$, are usually called Power Spectral Density of the detector (PSD) and represent the auto-correlation functions of the detector noise. The elements outside the diagonal, i.e., those with $I \neq J$, are usually referred to as the Cross Power Spectral Density (CSD) and quantify the noise between different detectors. The CSD is in principle a complex quantity that contains information about both the amplitude and phase of the noise correlation. This property is particularly useful for identifying and understanding common noise sources that affect multiple detectors.
In reality, both the PSD and the CSD will vary over time and, as a consequence, affect the stationarity of the noise. Here we ignore this complication, as we deal with simulated data. 

We refer to the PSD as $N_d(f)$ (diagonal) and to the CSD as $N_o(f)$ (off-diagonal). We assume that, in principle, every pair of detectors is correlated. Thus, for the triangular configuration, we express the covariance matrix of the noise as
\begin{align}
    N_{IJ}(f) &\equiv \begin{pmatrix}
        N_d(f) & N_o(f) & N_o(f) \\
        N_o(f) & N_d(f) & N_o(f) \\
        N_o(f) & N_o(f) & N_d(f)
    \end{pmatrix} \, ,
    \label{noisematrix}
\end{align}
where we assume that all auto-correlation channels have the same PSD and cross-correlation channels have the same CSD respectively, since the three interferometers have identical properties~\cite{Flauger:2020qyi}. We also assume that the geophysical environment at the three vertices is the same. 
This is indeed an assumption, as, in general, the PSDs of the three ET detectors could differ, as well as the noise correlations between the channels. This assumption is reasonable as a first approximation for the scope of this study, but we plan to relax it in future work.  There are already a few studies that account for unequal PSDs and introduce formalisms to estimate the noise in such cases (see \cite{Hartwig:2023pft, Kume:2024sbu, Janssens:2022cty, Liu:2024jna}). The positive definiteness of the covariance matrix of the noise implies that all its eigenvalues are real and non-negative, which gives the condition $-1/2\leq N_o(f)/N_d(f) \leq 1$ in each frequency bin. 

Following the results found by ~\cite{Janssens:2022tdj}, the NN is expected to have a declining behaviour as a function of frequency (in $\Omega_{\rm GW}$ units). For this reason, we use the following correlated noise template
\begin{align}
N_o(f) =   N_d(2.75{\, \rm Hz})\, r\left(\frac{f}{2.75\, \rm Hz}\right)^{n_{\rm noise}} \,,
\label{eq:correlated_noise_parametrization}
\end{align}
where $N_d(2.75{\, \rm Hz})$ is the amplitude of the noise at the pivot frequency $2.75\, \rm Hz$, while $r$ is the correlation of the noise at $2.75\, \rm Hz$ and $n_{\rm noise}$ is the spectral tilt of the noise. This template is justified by the results found by~\cite{Janssens:2022xmo, Saulson:1984yg, Badaracco:2019vjq, Janssens:2023anf, Amann:2020jgo}. 

In general, the noise CSD, $N_o$, could contain a complex phase~\cite{Liu:2024jna}, which induces different correlations between the real and imaginary parts of $\tilde{n}_{I}$ and $\tilde{n}_J$. However, such contributions are not expected to affect the reconstruction of the SGWB, since any complex phase in the noise can be reabsorbed in the definition of the stochastic signal, which is characterized just by its (real) covariance. Therefore, we will only consider the case where the CSD is real, although a complex phase could have influence the parameter estimation of resolved sources~\cite{Cireddu:2023ssf,Wong:2024hes} and, as a consequence, the residual SGWB. In Appendix~\ref{app:Noise_Phase}, we provide a proof that the a complex phase in the CSD would not impact the reconstruction of the parameters of the SGWB. The assumption that the covariance matrix of the noise is real corresponds to the case in which the complex phase is 0 (positive correlation) or $\pi$ (negative correlation). Although the reconstruction of the parameters of the SGWB is expected to be the same for these two cases, we perform the Bayesian analysis estimating also the sign of the (real) correlation, in order to show the capability of reconstructing faithfully also the noise, once an accurate template for the frequency dependence is provided. As already stressed, such a reconstruction of the noise will play a crucial role for the case of the resolved sources as in~\cite{Cireddu:2023ssf,Wong:2024hes}. For simplicity, we restrict the phase to either $0$ or $\pi$, leaving the discussion of a Bayesian analysis of the SGWB and noise with a complex CSD to future work.

\subsection{Characterization of the noise for the 2L configuration}
\label{sec: Characterization of the noise for the 2L configuration}
In the case of ET in the 2L configuration, we will measure data from two independent detectors. Here we assume that the noises in the 2L interferometers are uncorrelated, therefore we set the CSD to zero. This can be done since the detectors are far apart and we do not consider the impact of correlated magnetic noise, as mentioned before. 
In the case of uncorrelated detectors, the covariance matrix of the noise, computed by using Eq.~\eqref{def:PSD_noise}, reduces to the diagonal matrix
\begin{align}
   N_{IJ}(f) &\equiv \begin{pmatrix}
        N(f) & 0 \\
        0 & N(f) 
    \end{pmatrix} \, .
    \label{noisematrix_2L}
\end{align}
As before, we assume that the two detectors have identical PSDs, justified by their identical design and properties \cite{Branchesi:2023mws}.

\section{Bayesian analysis for the SGWB}

\label{sec:bayesian analyis}

\subsection{Likelihood for correlated and uncorrelated configurations}

\label{subsec:likelihood}

In GW experiments, we can perform the Fourier transform of the data over different time segments of duration $T_{\rm seg}$ for each detector channel, denoted as $\{\tilde{s}_I(t,f)\}$. The times $t$ are equally spaced with intervals of duration $T_{\rm seg}$, ranging from 0 to the total observation time $T_{\rm obs}$. In our analysis, we assume that all the signals from the resolved sources have been subtracted, thus the data consist only of the stochastic background(s) and noise. In a single time segment $t$ of duration $T_{\rm seg}=4\,s$, the SGWB, if of astrophysical origin, could be non-stationary due to the low number of sources contributing to the signal over such a short period. Moreover, the noise may contain multiple non-Gaussian contributions, due to glitches, seismic vibrations and scattering noise~\cite{LIGOScientific:2016gtq, aLIGO:2020wna, LIGOScientific:2014qfs, LIGO:2024kkz}. Furthermore, the PSD of the noise introduced in Section~\ref{section:ET configurations and noise model} could vary over time~\cite{Cornish:2020odn}. For these reasons, we perform a Bayesian search for a stochastic background by introducing a time-averaged estimator of a quadratic combination of the data, as in~\cite{LIGOScientific:2019vic, KAGRA:2021kbb}. For two channels or detectors $I$ and $J$, the estimator used in SGWB analysis is
\begin{equation}
    \hat{C}_{IJ}(f) \equiv \sum_t\frac{2}{T_{\rm{obs}} S_{0}(f)}\Re[\tilde{s}_{I}(t,f)\tilde{s}^{*}_{J}(t,f)]\,, 
    \label{eq:hatC}
\end{equation}
with the factor $S_{0}(f) \equiv 3H_{0}^{2}/10 \pi^{2} f^{3}$ used to match the units of the square of the data to those of the spectral energy density of the SGWB introduced in Eq.~\eqref{eq:twopoint_corr}.
Averaging over many time segments allows us to treat the estimator of the SGWB as a Gaussian random variable, due to the central limit theorem \cite{Romano:2016dpx}. In Section~\ref{sec:Parameter estimation outcomes}, we discuss in more detail the choice of $T_{\rm obs}$ that justifies this assumption. According to Eqs.~\eqref{eq:2pt} and ~\eqref{def:PSD_noise}, the average of the estimator of the SGWB is 
\begin{equation}
    \bar{C}_{IJ}(f) \equiv \left\langle \hat{C}_{IJ}(f)\right\rangle = \gamma_{IJ}(f)\Omega_{\rm GW}(f)+\frac{N_{IJ}(f)}{S_{0}(f)} \, ,
    \label{eq:barCIJ}
\end{equation}
where $\gamma_{IJ}$ is the normalized ORF. When the noise of two interferometers are uncorrelated, as in the LIGO-Virgo-KAGRA (LVK) case or ET in the 2L configuration, the estimator introduced in Eq.~\eqref{eq:hatC} is an unbiased estimator for the energy density of the SGWB \cite{LIGOScientific:2019vic}. In contrast, when noise correlation is present, as in ET in the triangular configuration, the correlation could introduce a non-negligible bias to the estimated amplitude and tilt of the SGWB. In~\cite{Romano:2016dpx} it has been shown that the variance of the estimator is proportional to the SGWB signal itself (intrinsic variance) and to the PSD of the noise,
\begin{equation}
    \Sigma_{IJ}(f) \equiv \left\langle \left[\hat{C}_{IJ}(f)-\bar{C}_{IJ}(f)\right]^2 \right\rangle = \frac{\bar{C}_{II}\bar{C}_{JJ}+\bar{C}_{IJ}^2}{2 N_{\textrm{seg}}} \, , 
    \label{eq:SigmaIJ}
\end{equation}
with $N_{\rm seg}\equiv T_{\rm obs}/T_{\rm seg}$ the number of time segments used in the analysis. 
The full likelihood of the data incorporates information from all $I$ and $J$ channels in the detector network. However, in our analysis, we focus only on the joint estimation of the SGWB parameters and, in the case of ET in the triangular configuration, the correlated noise parameters, since we aim to show that the parameters of the SGWB can be measured with sufficient precision even when $N_{IJ}(f)\neq 0$ (i.e., noise correlation is present)\footnote{Furthermore, the estimation of the auto-PSD of the noise requires a parametrization of all the contributions to the noise, which we do not consider here.}. We exclude from the likelihood the channels expected to provide most of the information about the noise auto-spectrum. Additionally, we fix the PSD to the reference value provided in~\cite{Branchesi:2023mws}. These choices, in principle, should give to us conservative estimates. For ET in the triangular configuration, we diagonalize the PSD, using the AET basis~\cite{Hogan:2001jn, Adams:2010vc,Flauger:2020qyi}. In this basis, it is typically assumed that the T channel is used to estimate the noise, as it is almost insensitive to the presence of a SGWB~\cite{Armstrong_1999}. Details on the transformation from the XYZ to the AET basis are given in Appendix~\ref{app:aet_xyz_bases}. Similarly, for the 2L configuration, we consider only one channel, namely the cross-correlation between the two detectors, as we assume that the auto-correlations are used to reconstruct the PSD. We can thus write down a Gaussian likelihood in a form that includes both configurations
\begin{equation}
    \mathcal{L} = \prod_{I,J} \frac{1}{\sqrt{2\pi \Sigma_{IJ}}} {\rm exp}\left[-\frac{1}{2}\frac{(\hat{C}_{IJ}-\bar{C}_{IJ})^2}{\Sigma_{IJ}}\right] \, , 
    \label{eq:likelihood_estimator}
\end{equation}
where, in the case of ET in the triangular configuration, we consider the pairs $(I,J)=\{(A,A),(E,E)\}$, while for the 2L configuration, we use only $(I,J)=(\rm ET1,ET2)$. In Appendix~\ref{app:Equivalence of different likelihoods}, we derive the Gaussian likelihood of the estimator, starting from the likelihood of the data, showing the consistency of our computations by using different likelihoods.

\subsection{Data Generation} 
\label{subsec:data_generation}

In this work, the signal and the noise are considered as Gaussian random variables with zero mean and covariance given by the PSD of the SGWB and noise respectively. As stressed in the previous subsection, although this is not expected to be true in a single time segment (at least for some astrophysical sources) and noise artifacts, the Gaussian approximation is valid once we average the data over a large number of time segments. Therefore, our analysis will not be altered by non-Gaussianities in the single time segments of zero average, which will be then neglected in the data generation. Thanks to the stationarity of the background and the noise, different frequencies are uncorrelated, therefore we can sample the data in each bin from $(f_{\rm max}-f_{\rm min})/\Delta f$ independent Gaussian distributions. In a similar fashion, the noise and the SGWB in different time segments are sampled from $N_{\rm seg}$ independent Gaussian distributions. Conversely, the data in different channels are correlated, implying they are generated from a multivariate Gaussian distribution. More specifically, in a time segment $t$ and a frequency bin $f$, we write the data as
\begin{equation}
    \begin{split}
        h_I(f) =& \frac{\xi_{I,\mathcal{R}}^{(h)}(f)+i\xi_{I,\mathcal{I}}^{(h)}(f)}{\sqrt{2}}\, \sqrt{\frac{T_{\rm seg}}{2}S_0(f)\Omega_{\rm GW}(f)}\,  \, , \\
        n_I(t,f) = & \frac{\xi_{I,\mathcal{R}}^{(n)}(t,f)+i\xi_{I,\mathcal{I}}^{(n)}(t,f)}{\sqrt{2}}\,\sqrt{\frac{T_{\rm seg}}{2}N_d(f)} \, ,
        \label{eq:h_n_generators}
    \end{split}
\end{equation}
where the generators of the signal and noise,  $\xi^{(h)}$ and $\xi^{(n)}$, are Gaussian random variables with zero mean and covariances equal to $\gamma_{IJ}(f)$ and $N_{IJ}(f)/N_d(f)$:
\begin{equation}
    \begin{split}
        & \left\langle \xi_{I,\mathcal{R}}^{(h)}(f)\right\rangle = \left\langle \xi_{I,\mathcal{I}}^{(h)}(f)\right\rangle =  \left\langle \xi_{I,\mathcal{R}}^{(n)}(f)\right\rangle = \left\langle \xi_{I,\mathcal{I}}^{(n)}(f)\right\rangle = 0 \, , \\
        &\left\langle \xi_{I,\mathcal{R}}^{(h)}(f)\xi_{J,\mathcal{R}}^{(h)}(f)\right\rangle = \left\langle \xi_{I,\mathcal{I}}^{(h)}(f)\xi_{J,\mathcal{I}}^{(h)}(f)\right\rangle = \gamma_{IJ}(f) \, , \\
        & \left\langle \xi_{I,\mathcal{R}}^{(n)}(f)\xi_{J,\mathcal{R}}^{(n)}(f)\right\rangle = \left\langle \xi_{I,\mathcal{I}}^{(n)}(f)\xi_{J,\mathcal{I}}^{(n)}(f)\right\rangle = \frac{N_{IJ}(f)}{N_d(f)} \, .
    \end{split}
\end{equation}
The real and imaginary parts of the generators are drawn from two identical yet independent multivariate Gaussians, sharing the same mean and covariances. Therefore, $\langle \xi_{I,\mathcal{R}}^{(h/n)}\xi_{J,\mathcal{I}}^{(h/n)}\rangle=0$. For the ET in the 2L configuration, the noise is uncorrelated, making our sampling approach equivalent to drawing the generators from two independent Gaussian distributions in the detector space. To be consistent with the likelihood given in Eq.~\eqref{eq:likelihood_estimator}, the data for ET in the triangular configuration are rotated in the AET basis according to the transformations described by Eq.~\eqref{eq:xyz_to_aet_rotation}.

\section{Parameter estimation outcomes}
\label{sec:Parameter estimation outcomes}

As discussed in the previous sections, we address correlated noise in the triangular configuration by modeling its frequency spectrum and reconstructing it alongside the SGWB spectrum through a Bayesian analysis. Specifically, since both the SGWB and the correlated noise spectra are modeled as power laws, as reported in Eqs. \eqref{eq:sgwb_parametrization} and \eqref{eq:correlated_noise_parametrization}, we reconstruct the posterior distribution of four parameters: the SGWB amplitude at 25 Hz, $A_{\textrm{GW}}$; the SGWB spectral tilt, $n_{\textrm{GW}}$; the correlated noise amplitude, parametrized as the ratio of the CSD to the PSD at 2.75 Hz, $r$; and the correlated noise spectral tilt, $n_{\textrm{noise}}$. In the 2L configuration, instead, since we assume there is no correlated noise, we reconstruct only the parameters of the SGWB.

\begin{table*} [t!]
\centering
\begin{tabular}{l@{\hskip 1cm}l@{\hskip 1cm}l@{\hskip 1cm}l}
\toprule
\hline
Parameter & Definition & Prior range   \\
\hline
\midrule
$ A_{\textrm{GW}} $ & Amplitude of the SGWB energy density at 25 Hz & $[10^{-13}, 3.5\cdot10^{-9}]$  \\
$n_{\textrm{GW}}$ & Tilt of the SGWB energy density & $[-10, 10]$ \\
$r$ & Ratio of the correlated noise CSD and the PSD at 2.75 Hz & $[-0.5, 1]$  \\
$n_{\textrm{noise}}$ & Tilt of the correlated noise CSD & $[-10, -5]$  \\
\hline
\end{tabular}
\caption{Prior ranges for all parameters. The prior for $A_{\textrm{GW}}$ is log-uniform, while uniform priors are used for all other parameters.}
\label{tab:priors}
\end{table*}

To perform the parameter estimation, we implement our likelihood given in Eq. \eqref{eq:likelihood_estimator} in \texttt{Bilby} \cite{Ashton:2018jfp,Romero-Shaw:2020owr,Morisaki:2023kuq} and reconstruct the posterior using the \texttt{Dynesty} sampler \cite{Speagle:2019ivv}. We use uniform priors for all the parameters except for $A_{\textrm{GW}}$, for which we use a log-uniform prior. We make a conservative choice for the prior ranges, which we assume to be fairly broad for all parameters, as reported in Table \ref{tab:priors}.
For the SGWB amplitude, the range is limited by the smallest value detectable by ET \cite{Branchesi:2023mws} and the current LVK upper bound \cite{KAGRA:2021kbb}. For the SGWB tilt, we select a broad range to ensure our results are not biased by the prior choice. Concerning the correlation parameter, its value at the pivot frequency is constrained to stay between -0.5 and 1, because of the semi-positive definiteness of the covariance matrix of the noise\footnote{By definition, actually, the noise CSD should be smaller than 1 and larger than -0.5 than the PSD at \textit{all} frequencies. However, there are some combinations of the prior values for $r$ and $n_{\textrm{noise}}$ that do not satisfy this requirement for our choice of the pivot frequency, 2.75 Hz. Therefore, we include in our implementation of the likelihood a constraint function to ensure that the sampler explores only regions of the parameter space that satisfy this requirement.}. 
Therefore, we adopt a uniform prior that spans all allowed real values within the range of -0.5 to 1. As discussed in Section \ref{section:ET configurations and noise model}, the CSD of the noise is complex in principle. However, since the complex phase can be reabsorbed in the definition of the SGWB without any impact on the reconstruction of the SGWB parameters, for illustrative purposes we consider only real values for the correlation parameter $r$, which corresponds to the sub-case of the phase which could assume values equal to 0 and $\pi$. A straightforward extension of this assumption would require a joint estimate of the real and imaginary parts of the correlation and we leave it for a future work. Finally, for the correlated noise tilt, we consider a relatively broad range centered around the value expected for Newtonian noise, which is $n_{\rm noise} =-8$ \cite{Saulson:1984yg,Badaracco:2019vjq,Amann:2020jgo,Janssens:2022xmo,Janssens:2024jln}.

We generate data for a total observation time of $T_{\textrm{obs}} = 1$ day, divided into time segments of $T_{\textrm{seg}} = 4$ s. We choose such a total observation time to guarantee that the averaged likelihood is Gaussian\footnote{We have verified that this holds for our simulated data. If the SGWB originates from stellar compact binary coalescences, for a large number of events this holds also for real data.}.
The data are generated in the frequency domain, as discussed in Section \ref{subsec:data_generation}, spanning the range from 1 Hz to 200 Hz, with a frequency resolution of $\Delta f = 1/T_{\textrm{seg}} = 0.25$ Hz. We do not extend the analysis to higher frequencies because, at 200 Hz, the sensitivity of ET to the SGWB has already dropped by 2 to 3 orders of magnitude compared to its maximum sensitivity, which occurs at $\sim$10 Hz \cite{Branchesi:2023mws}. Therefore, considering higher frequencies would not significantly increase the signal-to-noise ratio (SNR) of the considered signals. Moreover, beyond a few hundred Hz, the power-law assumption no longer holds for the SGWB produced by stellar compact binary coalescences, which is expected to be the dominant contribution to the SGWB in the frequency range explored by ground-based interferometers. 
We use the following values for the injected parameters: $A_{\textrm{GW}}^{\textrm{inj}} = 10^{-9}$, $n_{\textrm{GW}}^{\textrm{inj}} = 2/3$, $r^{\textrm{inj}} = \{-0.4, -0.2, 0.0, 0.2, 0.4, 0.6, 0.8\}$, and $n_{\textrm{noise}}^{\textrm{inj}} = -8$. 
For the injected SGWB, we have selected typical values for the amplitude and spectral tilt that are expected for a population of stellar-origin compact binaries, based on state-of-the art astrophysical models (see, e.g., \cite{KAGRA:2021kbb, Branchesi:2023mws, Capurri:2021zli, Bellomo:2021mer, Boco:2019teq, Perigois:2021ovr, Dvorkin:2016wac, Cusin:2019jpv}). 
For the correlated noise, we use the spectral tilt predicted for the Newtonian noise due to seismic body waves \cite{Saulson:1984yg,Badaracco:2019vjq,Amann:2020jgo,Janssens:2022xmo,Janssens:2024jln} and a set of different amplitudes, both positive and negative, in order to study the impact of both the sign and the strength of the correlation on the parameter estimation.
Throughout this work, we adopt the standard flat $\Lambda$CDM cosmology with parameter values from the Planck 2018 legacy release \cite{Planck:2018vyg}, with Hubble rate today corresponding to $H_{0}= 67.4$ km s$^{-1}$ Mpc$^{-1}$.

\begin{figure}[t!]
    \centering
    \includegraphics[width =.49 \textwidth]{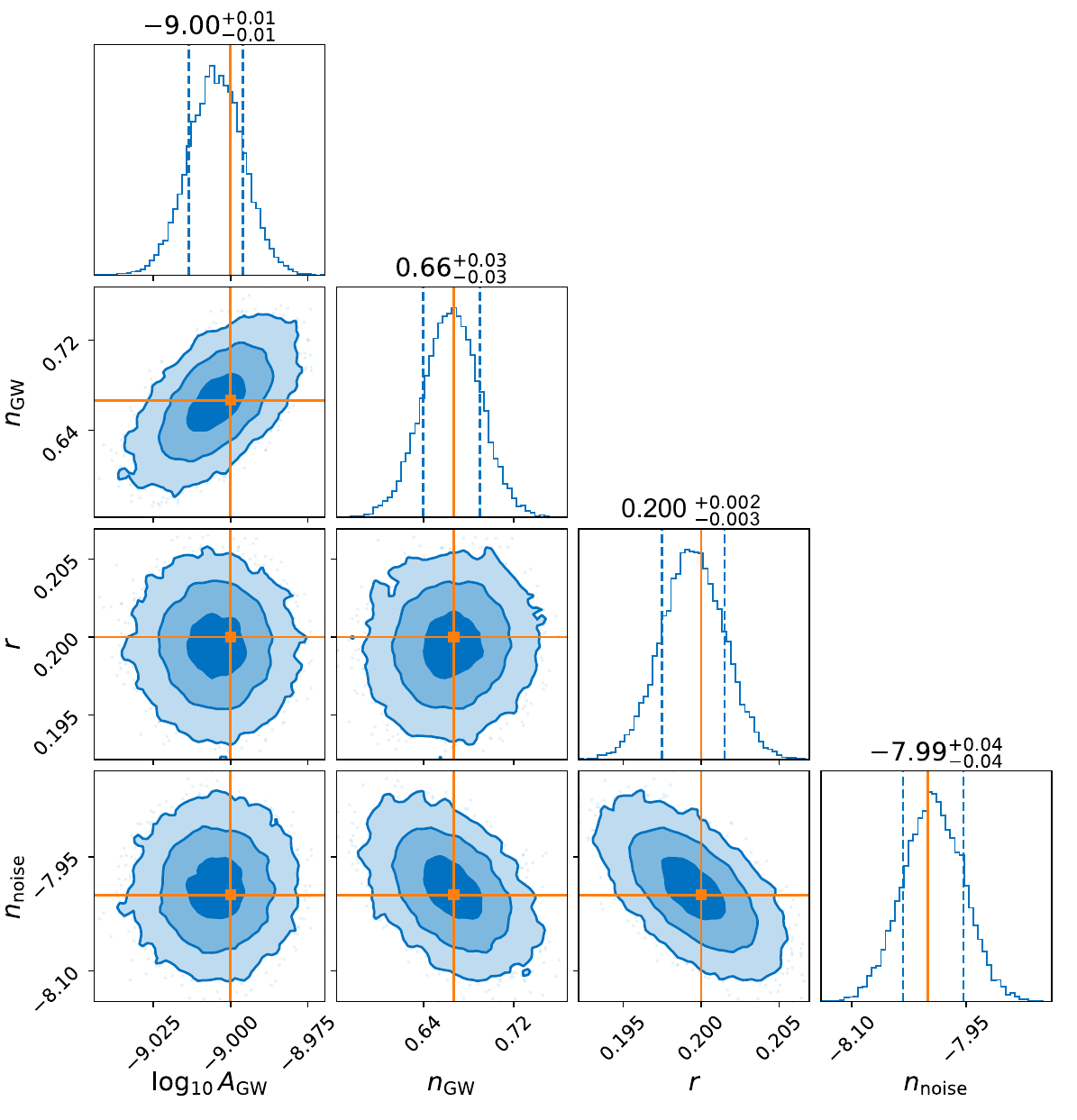}
    \caption{Corner plot of the posterior distributions for $\log_{10}A_\textrm{GW}$, $n_\textrm{GW}$, $r$, and $n_\textrm{noise}$. The shaded areas represent the 1-, 2-, and 3-$\sigma$ credible regions. The orange lines indicate the injected values for the four parameters: -9, 2/3, 0.2, and -8, respectively. } 
    \label{fig:corner_triangular}
\end{figure}

\begin{figure}[t!]
    \centering
    \includegraphics[width =.49 \textwidth]{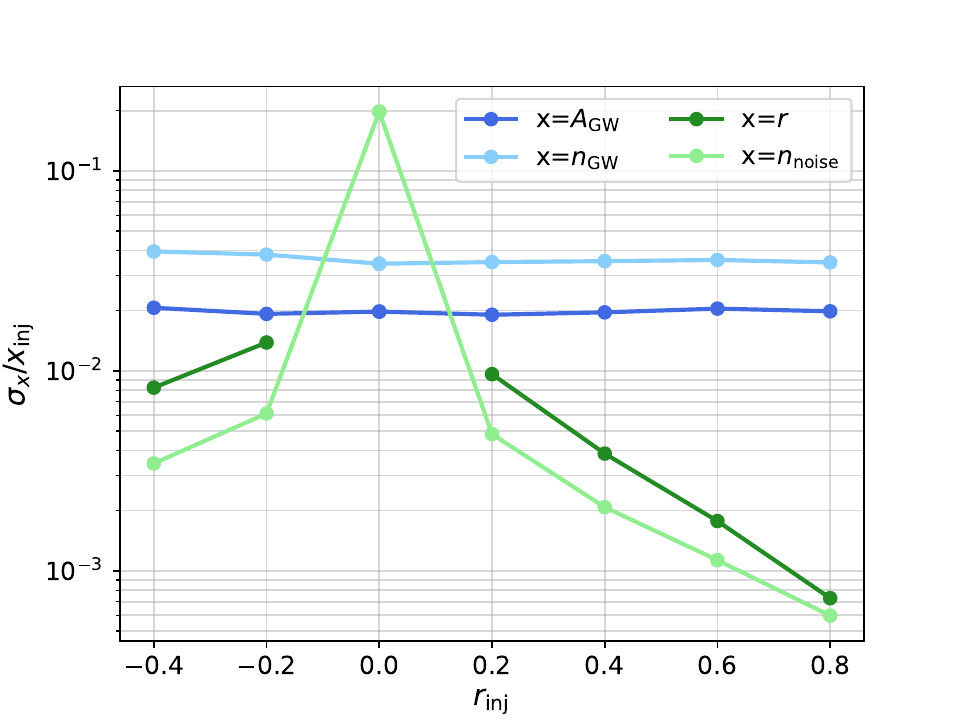}
    \caption{Relative error on the reconstructed parameters as a function of the injected noise correlation level for the triangular configuration.
    }
    \label{fig:relative_errors}
\end{figure}

Figure \ref{fig:corner_triangular} shows a corner plot of the reconstructed posterior distributions for the SGWB and the correlated noise parameters for ET in the triangular configuration. As an example, we show the plot for $r^{\textrm{inj}} = 0.2$. Remarkably, even with just one day of observations, all four parameters are reconstructed with high precision. This suggests that the presence of correlated noise does not hinder the accurate reconstruction of the SGWB parameters, provided that we correctly model its frequency dependence. This is a key assumption in our work, as it holds true in our case, where we deal with simulated data generated from the same noise model used for the PE, but it may not be valid for real data. Indeed, a realistic realization of Newtonian noise will be much more complex than a simple power law and will require more sophisticated models, which could be parametric or non-parametric (see, e.g., \cite{Martini:2024daa}), to be reconstructed and achieve a successful separation between signal and noise.  
As expected, we observe a positive correlation between the amplitude and the spectral tilt of the SGWB. This occurs because the chosen pivot frequency for the SGWB, 25 Hz, is higher than the frequencies at which ET is most sensitive to the SGWB ($\sim$ 10 Hz) \cite{Branchesi:2023mws}. As a consequence, a larger retrieved value of $n_{\textrm{GW}}$ requires a larger value of $A_{\textrm{GW}}$ to fit the injected signal, and vice versa. The amplitude and spectral tilt of the correlated noise are anti-correlated for the same reason. The only difference is that the chosen pivot frequency for the noise, 2.75 Hz, is now below 10 Hz.  Finally, we find a negative correlation between the spectral tilts of the SGWB and the noise. The reason is that if we retrieve a larger value of $n_{\textrm{GW}}$, we obtain less power at low frequencies, which requires a smaller (i.e., more negative) $n_{\textrm{noise}}$ to balance and fit the injected signal.

\begin{figure*}[t!]
    \centering
    \includegraphics[width =1 \textwidth]{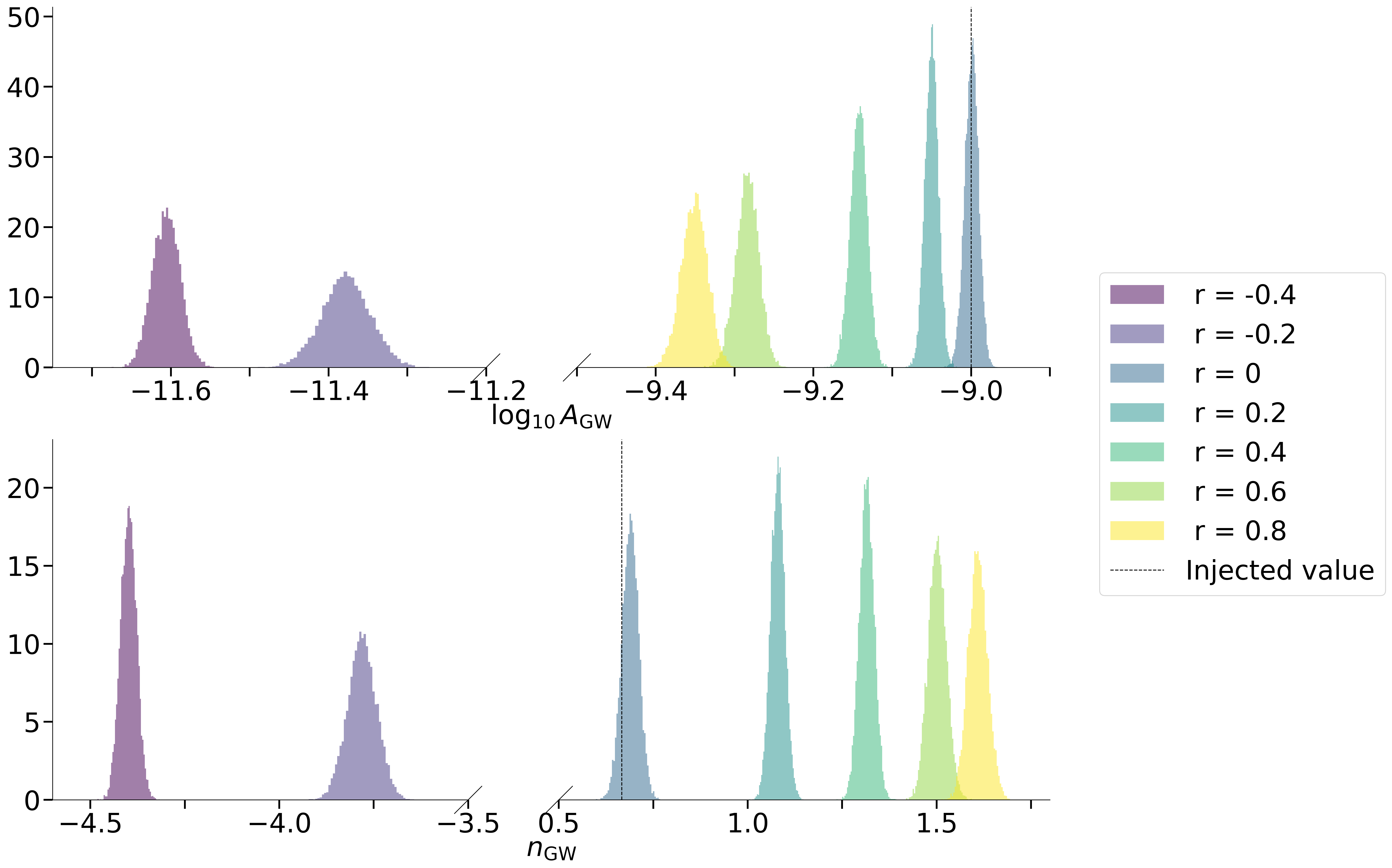}
    \caption{Posterior distribution of $\log_{10}A_{\rm GW}$ and $n_{\rm GW}$ for different values of $r$. Dashed vertical lines represent the true value of the injected parameters.}
    \label{fig:biasAGW}
\end{figure*}

We also investigate how different degrees of positive and negative correlation in the ET detector noise affect the measurement of the SGWB parameters. Figure \ref{fig:relative_errors} shows the relative errors of both the SGWB and noise parameters for different values of $r^{\textrm{inj}}$, spanning from -0.5 to 0.8. Remarkably, the errors on the SGWB parameters remain stable across this range, while the noise parameters become less accurately reconstructed as $r^{\textrm{inj}}$ approaches 0 \footnote{The absence of a central point in the curve for 
$r$ arises because the injected value is exactly zero, making the relative error undefined.}. This is consistent with the proof given in Appendix~\ref{app:Noise_Phase} that the reconstruction of the parameters of the SGWB is independent of the phase (and, therefore, on the sign) of the correlated noise. Furthermore, the reconstruction of the parameters of the noise is well understood: correlated noise with smaller power is inherently harder to reconstruct. We also checked that the relative error scales as the inverse of the SNR. For reference, the injected SGWB signal has SNR $\sim 67$ for every value of $r$, while the correlated noise with $r = 0.2$, if interpreted as a signal, has SNR $\sim 135$ \footnote{The SNR is evaluated according to the standard definition (see, e.g., Eq. A.11 of \cite{Branchesi:2023mws}), including a correlated noise contribution in the noise matrix.}.  In general, one would expect that different values of $r$ would result in a different sensitivity to the SGWB. Our finding that the SNR of the SGWB and, accordingly, the relative error on the reconstructed SGWB parameters do not change significantly with $r$,  may be attributed to the fact that, within our power-law model for the correlated noise, the range of frequencies where the CSD is comparable in magnitude to the PSD is very narrow and centered around a few Hz. Overall, our findings confirm that the presence of correlated noise does not hinder the measurement of SGWB parameters for any allowed value of $r$, provided its frequency spectrum is correctly modeled. 



To strengthen our argument, we perform a test to highlight the impact of neglecting correlated noise in SGWB analyses for the ET triangular configuration. As in the previous example, we generate different datasets with the same injected SGWB signal ($A_{\textrm{GW}}^{\textrm{inj}} = 10^{-9}$ and $n_{\textrm{GW}}^{\textrm{inj}} = 2/3$) and various correlated noise realizations ($r^{\textrm{inj}}$ ranging from -0.5 to 0.8, and $n_{\textrm{noise}}^{\textrm{inj}} = -8$). However, in this case, we do not perform the parameter estimation using the correct likelihood of Eq. \eqref{eq:likelihood_estimator}, but instead use a likelihood that does not account for the presence of correlated noise. This leads to strongly biased results, as shown in Figure \ref{fig:biasAGW}, where we plot the retrieved posterior distribution of the SGWB parameters for different values of $r^{\textrm{inj}}$.
As expected, the injected parameters are correctly retrieved only for $r^{\textrm{inj}} = 0$, since there is no correlated noise in this case and the likelihood that neglects it is accurate. For all other values of $r^{\textrm{inj}}$, the posteriors for both $A_{\textrm{GW}}$ and $n_{\textrm{GW}}$ are significantly biased, particularly for negative values of $r^{\textrm{inj}}$.

\begin{figure}[t!]
    \centering
    \includegraphics[width =.49 \textwidth]{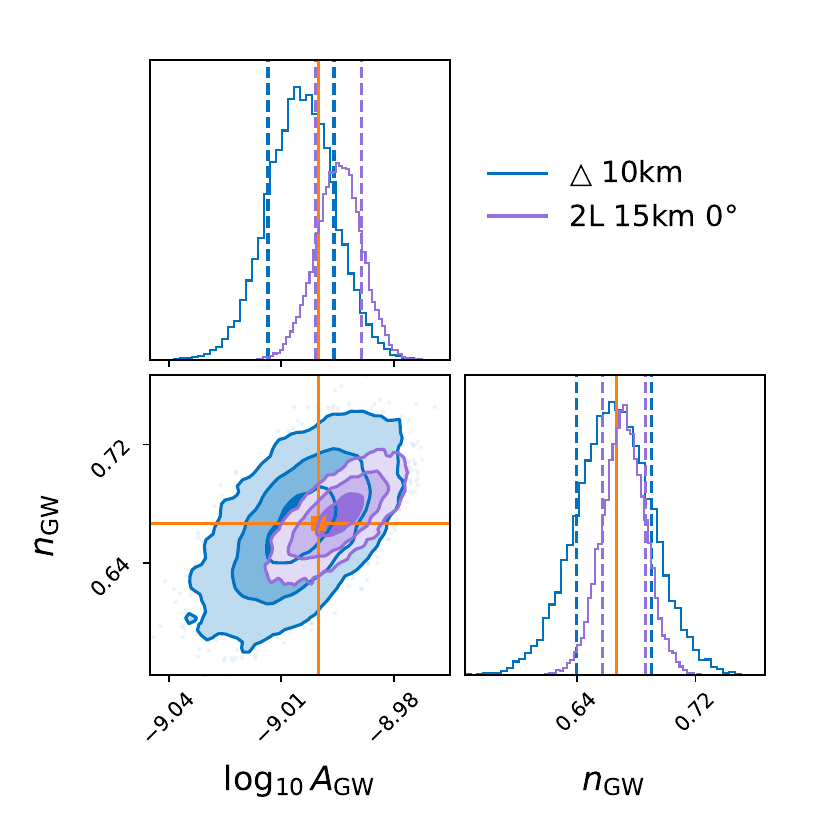}
    \caption{Comparison of the posteriors for the SGWB parameters obtained with ET in the triangular (blue) and 2L (purple) configurations. The orange lines represent the injected parameters, which are the same in both cases. The observing time is 1 day.}
    \label{fig:corner_triangular_2L}
\end{figure}

Finally, we compare the performance of the two proposed ET configurations: triangular and 2L. As commonly done in the literature, we compare the performance of a triangular ET with 10 km arms to that of two aligned L-shaped separated ET detectors with 15 km arms. 
In Figure \ref{fig:corner_triangular_2L}, we compare the posteriors for the SGWB parameters obtained with the two configurations. The observation time is one day in both cases, and the injected values for the SGWB parameters are the same. As expected, the parameter estimation yields unbiased results with both configurations. However, we find that the 2L performs slightly better, as the 1-$\sigma$ credible regions are approximately 1.5 times narrower than those in the triangle. This is mainly because the longer arms of the 2L provide better sensitivity to the SGWB compared to the triangle with 10 km arms. Moreover, the presence of correlated noise in the triangular configuration requires a parameter estimation with two additional parameters, which we marginalize over to obtain the posteriors on the SGWB parameters. The marginalization might result in further broadening of the posteriors.
We tested the statistical robustness of our Bayesian framework for both the triangular and the 2L configurations by producing the probability-probability (PP) plot for 100 injected signal and noise realizations, which we report in the Appendix.

\section{Conclusions and Outlook}

In this paper, we studied the impact of correlated noise on the detection and characterization of the SGWB with ET. Specifically, we introduced a new Bayesian framework to simultaneously reconstruct the parameters of the SGWB and the correlated noise, both modeled as power-laws in the frequency domain. 
Such an analysis is particularly relevant for the triangular configuration proposed for ET, which features a large nested triangular detector composed of three interferometers that share one arm each. In fact, correlated Newtonian noise generated by seismic body waves is expected to play a crucial role in this configuration, at low frequencies.

We derived a general likelihood for the SGWB that accounts for correlated noise and used it to perform parameter estimation on different sets of simulated data containing both an SGWB signal and a correlated noise contribution. 
Using an injected signal representative of a typical SGWB expected from stellar compact binary coalescences, along with an injected correlated noise characterized by the slope expected from Newtonian noise and varying amplitudes, we found that even with just one day of observation, we can reconstruct both the SGWB and noise parameters with percent-level accuracy.
As expected, the precision with which the noise parameters are reconstructed decreases as the noise amplitude approaches zero. Remarkably, however, the accuracy of the SGWB parameters remains stable across different values of the correlated noise amplitude. These findings indicate that it is possible to successfully detect and characterize the SGWB with ET in its triangular configuration, despite the presence of correlated noise, provided that the noise CSD is correctly modeled.

We further validated our results by analyzing the same simulated data using a likelihood that accounts only for the SGWB signal, without including the correlated noise. We found that the reconstructed SGWB parameters are significantly biased with respect to the injected values for every correlated noise amplitude used to generate the data. This result shows that, to successfully reconstruct the SGWB parameters, it is essential to include correlated noise in the Bayesian framework, when it is present.

Finally, for completeness, we compare the performance of the triangular configuration of ET with that of the other proposed configuration, which consists of two L-shaped detectors located at different sites in Europe. As it is usually done in the literature, we compare the triangular configuration with 10 km-long arms to the 2L configuration with aligned detectors and 15 km-long arms. 
We found that the 2L configuration outperforms the triangular one, with credible regions for the SGWB parameters that are roughly 1.5 times smaller. This slight improvement is mainly attributed to the longer interferometer arms, which enhance sensitivity to the SGWB. We tested the robustness of our Bayesian analysis using PP plots from 100 data realizations with parameters drawn from priors. For both the triangular and 2L configurations, all the distributions lie along the diagonal, mostly within 2-$\sigma$, showing that our reconstructed posteriors are statistically unbiased.

In conclusion, our analysis shows that it is possible to correctly reconstruct the parameters characterizing the SGWB even in the presence of correlated noise, provided that we have a valuable model for it. This is an assumption that surely holds in our analysis, as we deal with data simulated according to the same power-law noise model used for the analysis. When dealing with real data, more sophisticated noise models, possibly informed by seismograph data taken on-site, are needed to ensure the required level of precision in noise reconstruction. Indeed, we expect Newtonian noise due to seismic body waves to be the dominant source of correlated noise for ET, but the noise model can be further improved as other, less relevant sources of correlated noise are accurately modeled and eventually measured once the instrument becomes operative.  \\
Our analysis has been performed assuming a power law behaviour of the signal and noise. A natural extension would be to consider more sophisticated models for the correlated noise, as well as to consider SGWB sources which have a spectrum beyond a simple power law, such as phase transitions, or cosmic strings. Or to extend our Bayesian framework to perform a joint analyses of SGWB and resolved events.
\vspace{.5cm}
\paragraph*{Acknowledgments.}
\noindent
We thank N. Christensen, F. Cireddu, F. Fidecaro, J. Harms, M. Maggiore and R. Meyer for useful discussions and feedback on the draft. L.~V. acknowledges financial support from the project “LISA Global Fit’' funded by the ASI/Università di Trento Grant No. 2024-36-HH.0 - CUP F63C24000390001.
G.C., W.D.P. and A.R. acknowledge financial support from the project BIGA - ``Boosting Inference for Gravitational-wave Astrophysics" funded by the MUR Progetti di Ricerca di Rilevante Interesse Nazionale (PRIN) Bando 2022 - grant 20228TLHPE - CUP I53D23000630006.

\appendix

\section{Connection between XYZ and AET bases}
\label{app:aet_xyz_bases}

For detectors which have a triangular configuration, like ET, LISA, and Taiji, it is useful to work in the AET basis (if they have equal arm lengths and same noise levels), in which the covariance matrix of the noise is diagonal. Following~\cite{Flauger:2020qyi}, when the noise of the three interferomers have identical properties, the matrix of change of basis from XYZ to AET is
\begin{equation}
    R=
    \begin{pmatrix}
        -\frac{1}{\sqrt{2}} & 0 & \frac{1}{\sqrt{2}} \\
        \frac{1}{\sqrt{6}}  & -\frac{2}{\sqrt{6}}  & \frac{1}{\sqrt{6}}  \\
        \frac{1}{\sqrt{3}} & \frac{1}{\sqrt{3}} & \frac{1}{\sqrt{3}}
    \end{pmatrix} \, .
    \label{eq:xyz_to_aet_rotation}
\end{equation}
The data in the AET basis can be written as a function of the data in the XYZ basis as 
\begin{equation}
    \begin{split}
        \tilde{s}_A(f) =& 
        \frac{\tilde{s}_Z(f)-\tilde{s}_X(f)}{\sqrt{2}}  \, , \\
        \tilde{s}_E(f) =& \frac{\tilde{s}_X(f)-2\tilde{s}_Y(f)+\tilde{s}_Z(f)}{\sqrt{6}} \, , \\
        \tilde{s}_T(f)=&\frac{\tilde{s}_X(f)+\tilde{s}_Y(f)+\tilde{s}_Z(f)}{\sqrt{3}} \,   \, . 
    \end{split} 
\end{equation}    
The (diagonal) covariance matrix of the noise in the AET basis is 
\begin{equation}
	N_{IJ} \equiv \begin{pmatrix}
	N_d-N_o & 0 & 0 \\0 &  N_d-N_o & 0 \\ 0 & 0 & N_d+2N_o
	\end{pmatrix} \, ,
\end{equation}
with $N_o$ parametrized as in Eq.~\eqref{eq:correlated_noise_parametrization}. In a similar way, if the ORFs have identical properties for the three channels, the ORF matrix is also diagonal in the AET basis,  
\begin{equation}
	\gamma_{IJ} \equiv \begin{pmatrix}
	\gamma_d-\gamma_o & 0 & 0 \\0 &  \gamma_d-\gamma_o & 0 \\ 0 & 0 & \gamma_d+2\gamma_o
	\end{pmatrix} \, ,
\end{equation}
where $\gamma_d$ and $\gamma_o$ are the $\gamma_{IJ}$ defined in Eq.~\eqref{eq:orf_general_basis} when $I=J$ and $I\neq J$. The ORFs have been computed by using the code \texttt{GWBird}~\cite{GWBird}.

In the AET basis, with a spectral matrix as the one defined in Eq.~\eqref{noisematrix}, the data can be therefore considered as three uncorrelated Gaussian random variables of zero mean and covariance determined by the values of the PSD, CSD and ORF in the XYZ basis. As shown in~\cite{Flauger:2020qyi}, for the triangular detectors the $T$ (null) channel is (almost) insensitive to the SGWB (for equal arm length and noise levels), therefore it is possible to exploit this channel to estimate just  the PSD. As stated in the main text, because of this reason, in our analysis we do not consider the $T$ channel in the Bayesian analysis, assuming that it provides an exact information on the PSD of the noise. Furthermore, from the expression of the covariance of the data in the AET basis, it is clear that, in order to have a positive-definite covariance matrix, we require that all the eigenvalues are real. For the covariance of the SGWB this is true by definition, because $-1/2 \leq \gamma_o(f)/\gamma_d(f)\leq 1$. For the PSD and CSD of the noise, the condition $-1/2\leq N_o(f)/N_d(f)\leq 1$, corresponds to a constraint on the prior of the parameters of the correlated noise introduced in Eq.~\eqref{eq:correlated_noise_parametrization}.

\section{Insensitivity of SGWB parameters reconstruction to the complex phase of the noise}
\label{app:Noise_Phase}

In general, the covariance matrix of the noise written in Eq.~\eqref{noisematrix} could assume complex values in the elements outside the diagonal. The only constraints on the covariance matrix of the noise are indeed that the noise matrix is Hermitian, $N_{IJ}^*(f) = N_{JI}(f)$, and positive definite, i.e., all the eigenvalues are real and non-negative. It is possible therefore to parametrize it as
\begin{equation}
    N_{IJ}(f) =e^{i[\psi_I(f)-\psi_J(f)]}\left[ N_d(f)\delta_{IJ}+(1-\delta_{IJ})| N_o(f)|\right] \, , 
\end{equation}
where the combination $\psi_I(f)-\psi_J(f)$ sets the sign (and eventually the imaginary part) of the correlation, while $|N_o(f)|$ its amplitude. We note then that the SGWB has zero mean, (i.e., $\langle h \rangle$=0) therefore the expectation value of its phase is unobservable. To see this explicitly, we define the ‘‘phase-shifted SGWB'' as 
\begin{equation}
    \tilde{h}_I^{\rm shifted}(f) \equiv \tilde{h}_I(f)e^{-i\psi_I(f)} \, . 
\end{equation}
If this SGWB has zero mean and covariance equal to Eq.~\eqref{eq:2pt}, it is undistinguishable from the SGWB described by $\tilde{h}_I(f)$. In other words, there is no observable feature that allows to discriminate between $\tilde{h}_I^{\rm shifted}(f)$ and $\tilde{h}_I(f)$. Therefore, it is possible to re-write the data by factorizing out the complex phase which characterizes the noise,
\begin{equation}
    \tilde{s}_I(f) = e^{i\psi_I(f)}\left[\tilde{h}_I^{\rm shifted}(f)+e^{-\psi_I(f)}\tilde{n}_I(f)\right] \, .
    \label{eq:data_factorized_phase}
\end{equation}
According to this parametrization, the covariance matrix can be written as the sum of the covariances given in Eqs.~\eqref{eq:2pt},~\eqref{def:PSD_noise}, where the PSD of the noise assumes now just real and positive values, times a global phase $e^{i[\psi_I(f)-\psi_J(f)]}$. In the case of a Gaussian likelihood, like the one introduced in Eq.~\eqref{eq:likelihood_data}, the global phase in the covariance matrix would give an additional constant factor in the entropy term (the determinant of the covariance), while in the exponential it simplifies with the external phase of the data which appears in Eq.~\eqref{eq:data_factorized_phase}. Therefore, we conclude that the likelihood of the data with a complex covariance matrix of the noise is equivalent to the likelihood of the data with a ‘‘phase-shifted SGWB'' and a covariance matrix of the noise with real and positive entries. Since the ‘‘phase-shifted SGWB'' is characterized by the same parameters of the ‘‘original SGWB'', we can estimate the amplitude and the tilt introduced in Eq.~\eqref{eq:sgwb_parametrization} by neglecting the phases $\psi_I(f)$ in the covariance matrix of the noise. 

We stress that the phase of the CSD can be factorized out in the likelihood just in the case in which only signals with zero mean are present. In the case of resolved sources, as the one discussed in~\cite{Cireddu:2023ssf,Wong:2024hes}, where the coherence of the GWs is a fundamental feature of the signal, a ‘‘phase-shifted GW'' is not equivalent to the ‘‘original GW''.

\section{Equivalence of different likelihoods}
\label{app:Equivalence of different likelihoods}

In this work, we consider the data distributed as complex Gaussian random variables of zero mean, therefore in a single time segment $t$ of duration $T_{\rm seg}$, their likelihood is
\begin{equation}
    \mathcal{L} = \prod_{t,f} \frac{1}{\pi^{N_{\rm det}}{\rm det}\left(\bar{C}_{IJ}\right)}e^{-\sum_{I,J}\tilde{s}_I^\dagger(t,f)\left(\bar{C}^{-1}(f)\right)_{\, IJ}\tilde{s}(t,f)} \, ,
    \label{eq:likelihood_data}
\end{equation}
where $N_{\rm det}$ is the number of channels in the detector network, while the covariance of the data has been defined in Eq.~\eqref{eq:barCIJ}. In~\cite{Romano:2016dpx,KAGRA:2021kbb, Christensen:2018iqi}, an estimator of the SGWB has been defined by taking a quadratic combination of the data,
\begin{equation}
    \hat{c}_{IJ}(t,f) \equiv \frac{2}{T_{\rm seg}S_0(f)}\Re[\tilde{s}_I(t,f)\tilde{s}^*_J(t,f)] \, . 
\end{equation}
The $\hat{c}_{IJ}$ matrix is the real part of the square of correlated complex Gaussian random variables, therefore it follows a Wishart distribution~\cite{1626a8bf-8479-36bf-b395-22e4def81091}, with dimensions equal to the number of channels in the network $N_{\rm det}$ and 2 d.o.f. (the real and imaginary parts of the data). In this case the Wishart is
\begin{equation}
    \mathcal{L} = \prod_{t,f}\frac{\left[{\rm det}\left(\hat{c}_{IJ}(t,f)\right)\right]^{(N_{\rm det}-1)/2}e^{-\sum_{IJ}\left(\bar{C}^{-1}(f)\right)_{IJ}\hat{c}_{IJ}(t,f)}}{\pi^{N_{\rm det}}\left[{\rm det}\left(\bar{C}_{IJ}(f)\right)\right]} \, .
\end{equation} 
In the case of ET in a triangular configuration, we work in the diagonal basis, therefore the Wishart distribution reduces to the sum of many independent $\chi^2$ distributions with two degrees of freedom (one for the real and one for the imaginary part of the data), which can be written as
\begin{equation}
    \mathcal{L} = \prod_{t,f,I} \frac{1}{\bar{C}_{II}(f)}{\rm exp}\left(-\frac{\hat{c}_{II}(t,f)}{\bar{C}_{II}(f)}\right) \, ,
    \label{eq:likelihood_estimator_timesegment_diagonal}
\end{equation}
with $I=\{A,E\}$. In the case of ET in the 2L configuration, we just focus on the cross-correlation between two different channels, therefore we would need to compute the marginal Wishart distribution for $\hat{c}_{IJ}$. However, in this case, the likelihood cannot be written in a simple analytic form and we do not consider it.

The distribution of the estimator of the SGWB defined in Eq.~\eqref{eq:hatC}, which is the average over many time segments of $\hat{c}_{IJ}(t,f)$, is distributed according to a Wishart of dimension $N_{\rm det}$ with $2 N_{\rm seg}$ d.o.f. Thanks to the central limit theorem, such a distribution is a multivariate Gaussian distribution of mean $\bar{C}_{IJ}$ and covariance $\Sigma_{IJ}$, defined in Eqs.~\eqref{eq:barCIJ},~\eqref{eq:SigmaIJ} respectively.

For ET in the triangular configuration, we have checked that the results of Figures~\ref{fig:relative_errors} are the same for the Bayesian analyses
 performed with the likelihoods~\eqref{eq:likelihood_data},~\eqref{eq:likelihood_estimator_timesegment_diagonal},~\eqref{eq:likelihood_estimator}.

\section{PP-plots}

\begin{figure}[t!]
    \centering
    \includegraphics[width =.49 \textwidth]{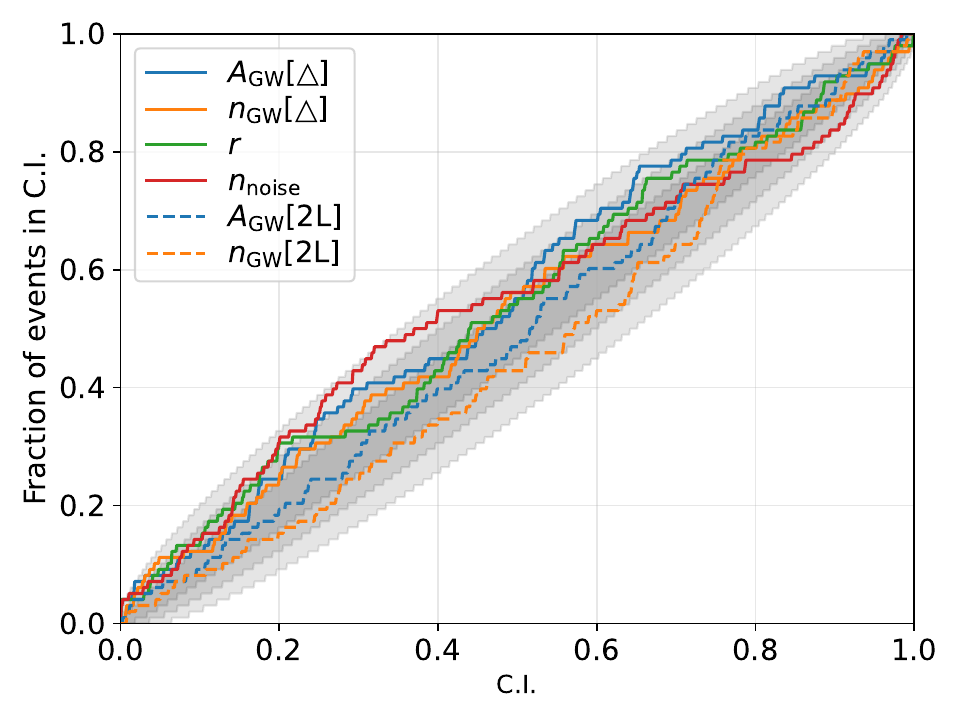}
    \caption{Probability-Probability (PP) plot for the posterior distributions of $A_\textrm{GW}$, $n_\textrm{GW}$, $r$, and $n_\textrm{noise}$, for the triangular (solid lines) and 2L (dashed lines) configurations. The shaded areas represent the 1-, 2-, and 3-$\sigma$ confidence levels due to finite sample size. }
    \label{fig:pp_plot}
\end{figure}

We test the statistical robustness of our Bayesian analysis by producing a probability-probability (PP) plot for all the reconstructed posteriors. First, we generate 100 different realizations of the data, randomly extracting the values of the SGWB and noise parameters from the priors reported in Table \ref{tab:priors}. We then calculate the fraction of injected realizations that fall within any confidence interval for the individual parameters $A_\textrm{GW}$, $n_\textrm{GW}$, $r$, and $n_\textrm{noise}$, for both the triangular and 2L configurations. The expected distribution is a diagonal line, with credible intervals computed from a beta distribution, as in \cite{Cameron:2010bh}. As shown in Figure \ref{fig:pp_plot}, all the PP plots lie along the diagonal within the 2-$\sigma$ credible interval in most cases. This confirms that the reconstructed posteriors are statistically unbiased.

\bibliographystyle{ieeetr}
\bibliography{Biblio_PRD.bib}

\begin{thebibliography}{100}

\bibitem{Punturo:2010zz}
M.~Punturo {\em et~al.}, ``{The Einstein Telescope: A third-generation
  gravitational wave observatory},'' {\em Class. Quant. Grav.}, vol.~27,
  p.~194002, 2010.

\bibitem{ET:2019dnz}
M.~Maggiore {\em et~al.}, ``{Science Case for the Einstein Telescope},'' {\em
  JCAP}, vol.~03, p.~050, 2020.

\bibitem{Regimbau:2011rp}
T.~Regimbau, ``{The astrophysical gravitational wave stochastic background},''
  {\em Res. Astron. Astrophys.}, vol.~11, pp.~369--390, 2011.

\bibitem{Caprini:2018mtu}
C.~Caprini and D.~G. Figueroa, ``{Cosmological Backgrounds of Gravitational
  Waves},'' {\em Class. Quant. Grav.}, vol.~35, no.~16, p.~163001, 2018.

\bibitem{Janssens:2022tdj}
K.~Janssens {\em et~al.}, ``{Correlated 1\textendash{}1000~Hz magnetic field
  fluctuations from lightning over Earth-scale distances and their impact on
  gravitational wave searches},'' {\em Phys. Rev. D}, vol.~107, no.~2,
  p.~022004, 2023.

\bibitem{Janssens:2023anf}
K.~Janssens, ``{Prospects for an isotropic gravitational wave background
  detection with Earth-based interferometric detectors and the threat of
  correlated noise},'' in {\em {57th Rencontres de Moriond on Gravitation}}, 5
  2023.

\bibitem{Janssens:2024jln}
K.~Janssens {\em et~al.}, ``{Correlated 0.01Hz-40Hz seismic and Newtonian noise
  and its impact on future gravitational-wave detectors},'' 2 2024.

\bibitem{di2021seismological}
M.~Di~Giovanni, C.~Giunchi, G.~Saccorotti, A.~Berbellini, L.~Boschi,
  M.~Olivieri, R.~De~Rosa, L.~Naticchioni, G.~Oggiano, M.~Carpinelli, {\em
  et~al.}, ``A seismological study of the sos enattos area—the sardinia
  candidate site for the einstein telescope,'' {\em Seismological Research
  Letters}, vol.~92, no.~1, pp.~352--364, 2021.

\bibitem{andric2020simulations}
T.~Andric and J.~Harms, ``Simulations of gravitoelastic correlations for the
  sardinian candidate site of the einstein telescope,'' {\em Journal of
  Geophysical Research: Solid Earth}, vol.~125, no.~10, p.~e2020JB020401, 2020.

\bibitem{ETdesignRep}
{ET Steering Committee Editorial Team}, ``{Collection of documents on the ET
  design study and science case},''

\bibitem{Branchesi:2023mws}
M.~Branchesi {\em et~al.}, ``{Science with the Einstein Telescope: a comparison
  of different designs},'' {\em JCAP}, vol.~07, p.~068, 2023.

\bibitem{Janssens:2022xmo}
K.~Janssens, G.~Boileau, N.~Christensen, F.~Badaracco, and N.~van Remortel,
  ``{Impact of correlated seismic and correlated Newtonian noise on the
  Einstein Telescope},'' {\em Phys. Rev. D}, vol.~106, no.~4, p.~042008, 2022.

\bibitem{Cireddu:2023ssf}
F.~Cireddu, M.~Wils, I.~C.~F. Wong, P.~T.~H. Pang, T.~G.~F. Li, and
  W.~Del~Pozzo, ``{Likelihood for a network of gravitational-wave detectors
  with correlated noise},'' {\em Phys. Rev. D}, vol.~110, no.~10, p.~104060,
  2024.

\bibitem{Romano:2016dpx}
J.~D. Romano and N.~J. Cornish, ``{Detection methods for stochastic
  gravitational-wave backgrounds: a unified treatment},'' {\em Living Rev.
  Rel.}, vol.~20, no.~1, p.~2, 2017.

\bibitem{Hogan:2001jn}
C.~J. Hogan and P.~L. Bender, ``{Estimating stochastic gravitational wave
  backgrounds with Sagnac calibration},'' {\em Phys. Rev. D}, vol.~64,
  p.~062002, 2001.

\bibitem{Adams:2010vc}
M.~R. Adams and N.~J. Cornish, ``{Discriminating between a Stochastic
  Gravitational Wave Background and Instrument Noise},'' {\em Phys. Rev. D},
  vol.~82, p.~022002, 2010.

\bibitem{Janssens:2022cty}
K.~Janssens, G.~Boileau, M.-A. Bizouard, N.~Christensen, T.~Regimbau, and
  N.~van Remortel, ``{Formalism for power spectral density estimation for
  non-identical and correlated noise using the null channel in Einstein
  Telescope},'' {\em Eur. Phys. J. Plus}, vol.~138, no.~4, p.~352, 2023.
\newblock [Erratum: Eur.Phys.J.Plus 138, 446 (2023)].

\bibitem{Allen:1997ad}
B.~Allen and J.~D. Romano, ``{Detecting a stochastic background of
  gravitational radiation: Signal processing strategies and sensitivities},''
  {\em Phys. Rev. D}, vol.~59, p.~102001, 1999.

\bibitem{Caprini:2019pxz}
C.~Caprini, D.~G. Figueroa, R.~Flauger, G.~Nardini, M.~Peloso, M.~Pieroni,
  A.~Ricciardone, and G.~Tasinato, ``{Reconstructing the spectral shape of a
  stochastic gravitational wave background with LISA},'' {\em JCAP}, vol.~11,
  p.~017, 2019.

\bibitem{Flauger:2020qyi}
R.~Flauger, N.~Karnesis, G.~Nardini, M.~Pieroni, A.~Ricciardone, and
  J.~Torrado, ``{Improved reconstruction of a stochastic gravitational wave
  background with LISA},'' {\em JCAP}, vol.~01, p.~059, 2021.

\bibitem{LISACosmologyWorkingGroup:2022jok}
P.~Auclair {\em et~al.}, ``{Cosmology with the Laser Interferometer Space
  Antenna},'' {\em Living Rev. Rel.}, vol.~26, no.~1, p.~5, 2023.

\bibitem{Braglia:2024kpo}
M.~Braglia {\em et~al.}, ``{Gravitational waves from inflation in LISA:
  reconstruction pipeline and physics interpretation},'' {\em JCAP}, vol.~11,
  p.~032, 2024.

\bibitem{Blanco-Pillado:2024aca}
J.~J. Blanco-Pillado, Y.~Cui, S.~Kuroyanagi, M.~Lewicki, G.~Nardini,
  M.~Pieroni, I.~Y. Rybak, L.~Sousa, and J.~M. Wachter, ``{Gravitational waves
  from cosmic strings in LISA: reconstruction pipeline and physics
  interpretation},'' 5 2024.

\bibitem{Caprini:2024hue}
C.~Caprini, R.~Jinno, M.~Lewicki, E.~Madge, M.~Merchand, G.~Nardini,
  M.~Pieroni, A.~Roper~Pol, and V.~Vaskonen, ``{Gravitational waves from
  first-order phase transitions in LISA: reconstruction pipeline and physics
  interpretation},'' {\em JCAP}, vol.~10, p.~020, 2024.

\bibitem{Phinney:2001di}
E.~S. Phinney, ``{A Practical theorem on gravitational wave backgrounds},'' 7
  2001.

\bibitem{ValbusaDallArmi:2023ydl}
L.~Valbusa~Dall'Armi, A.~Nishizawa, A.~Ricciardone, and S.~Matarrese,
  ``{Circular Polarization of the Astrophysical Gravitational Wave
  Background},'' {\em Phys. Rev. Lett.}, vol.~131, no.~4, p.~041401, 2023.

\bibitem{Belgacem:2024ohp}
E.~Belgacem, F.~Iacovelli, M.~Maggiore, M.~Mancarella, and N.~Muttoni, ``{The
  spectral density of astrophysical stochastic backgrounds},'' 11 2024.

\bibitem{Grishchuk:1974ny}
L.~P. Grishchuk, ``{Amplification of gravitational waves in an istropic
  universe},'' {\em Zh. Eksp. Teor. Fiz.}, vol.~67, pp.~825--838, 1974.

\bibitem{Starobinsky:1979ty}
A.~A. Starobinsky, ``{Spectrum of relict gravitational radiation and the early
  state of the universe},'' {\em JETP Lett.}, vol.~30, pp.~682--685, 1979.

\bibitem{Guth:1980zm}
A.~H. Guth, ``{The Inflationary Universe: A Possible Solution to the Horizon
  and Flatness Problems},'' {\em Phys. Rev. D}, vol.~23, pp.~347--356, 1981.

\bibitem{Starobinsky:1980te}
A.~A. Starobinsky, ``{A New Type of Isotropic Cosmological Models Without
  Singularity},'' {\em Phys. Lett. B}, vol.~91, pp.~99--102, 1980.

\bibitem{Linde:1981mu}
A.~D. Linde, ``{A New Inflationary Universe Scenario: A Possible Solution of
  the Horizon, Flatness, Homogeneity, Isotropy and Primordial Monopole
  Problems},'' {\em Phys. Lett. B}, vol.~108, pp.~389--393, 1982.

\bibitem{Albrecht:1982wi}
A.~Albrecht and P.~J. Steinhardt, ``{Cosmology for Grand Unified Theories with
  Radiatively Induced Symmetry Breaking},'' {\em Phys. Rev. Lett.}, vol.~48,
  pp.~1220--1223, 1982.

\bibitem{Guzzetti:2016mkm}
M.~C. Guzzetti, N.~Bartolo, M.~Liguori, and S.~Matarrese, ``{Gravitational
  waves from inflation},'' {\em Riv. Nuovo Cim.}, vol.~39, no.~9, pp.~399--495,
  2016.

\bibitem{Sorbo:2011rz}
L.~Sorbo, ``{Parity violation in the Cosmic Microwave Background from a
  pseudoscalar inflaton},'' {\em JCAP}, vol.~06, p.~003, 2011.

\bibitem{Badger:2024ekb}
C.~Badger, H.~Duval, T.~Fujita, S.~Kuroyanagi, A.~Romero-Rodr\'\i{}guez, and
  M.~Sakellariadou, ``{Detection prospects of gravitational waves from SU(2)
  axion inflation},'' {\em Phys. Rev. D}, vol.~110, no.~8, p.~084063, 2024.

\bibitem{Garcia-Bellido:2023ser}
J.~Garcia-Bellido, A.~Papageorgiou, M.~Peloso, and L.~Sorbo, ``{A flashing
  beacon in axion inflation: recurring bursts of gravitational waves in the
  strong backreaction regime},'' {\em JCAP}, vol.~01, p.~034, 2024.

\bibitem{Biagetti:2013kwa}
M.~Biagetti, M.~Fasiello, and A.~Riotto, ``{Enhancing Inflationary Tensor Modes
  through Spectator Fields},'' {\em Phys. Rev. D}, vol.~88, p.~103518, 2013.

\bibitem{Cook:2011hg}
J.~L. Cook and L.~Sorbo, ``{Particle production during inflation and
  gravitational waves detectable by ground-based interferometers},'' {\em Phys.
  Rev. D}, vol.~85, p.~023534, 2012.
\newblock [Erratum: Phys.Rev.D 86, 069901 (2012)].

\bibitem{Barnaby:2012xt}
N.~Barnaby, J.~Moxon, R.~Namba, M.~Peloso, G.~Shiu, and P.~Zhou, ``{Gravity
  waves and non-Gaussian features from particle production in a sector
  gravitationally coupled to the inflaton},'' {\em Phys. Rev. D}, vol.~86,
  p.~103508, 2012.

\bibitem{Bartolo:2015qvr}
N.~Bartolo, D.~Cannone, A.~Ricciardone, and G.~Tasinato, ``{Distinctive
  signatures of space-time diffeomorphism breaking in EFT of inflation},'' {\em
  JCAP}, vol.~03, p.~044, 2016.

\bibitem{Ricciardone:2016lym}
A.~Ricciardone and G.~Tasinato, ``{Primordial gravitational waves in supersolid
  inflation},'' {\em Phys. Rev. D}, vol.~96, no.~2, p.~023508, 2017.

\bibitem{Capurri:2020qgz}
G.~Capurri, N.~Bartolo, D.~Maino, and S.~Matarrese, ``{Let Effective Field
  Theory of Inflation flow: stochastic generation of models with red/blue
  tensor tilt},'' {\em JCAP}, vol.~11, p.~037, 2020.

\bibitem{Tomita:1975kj}
K.~Tomita, ``{Evolution of Irregularities in a Chaotic Early Universe},'' {\em
  Prog. Theor. Phys.}, vol.~54, p.~730, 1975.

\bibitem{Matarrese:1992rp}
S.~Matarrese, O.~Pantano, and D.~Saez, ``{A General relativistic approach to
  the nonlinear evolution of collisionless matter},'' {\em Phys. Rev. D},
  vol.~47, pp.~1311--1323, 1993.

\bibitem{Matarrese:1993zf}
S.~Matarrese, O.~Pantano, and D.~Saez, ``{General relativistic dynamics of
  irrotational dust: Cosmological implications},'' {\em Phys. Rev. Lett.},
  vol.~72, pp.~320--323, 1994.

\bibitem{Matarrese:1997ay}
S.~Matarrese, S.~Mollerach, and M.~Bruni, ``{Second order perturbations of the
  Einstein-de Sitter universe},'' {\em Phys. Rev. D}, vol.~58, p.~043504, 1998.

\bibitem{Acquaviva:2002ud}
V.~Acquaviva, N.~Bartolo, S.~Matarrese, and A.~Riotto, ``{Second order
  cosmological perturbations from inflation},'' {\em Nucl. Phys. B}, vol.~667,
  pp.~119--148, 2003.

\bibitem{Mollerach:2003nq}
S.~Mollerach, D.~Harari, and S.~Matarrese, ``{CMB polarization from secondary
  vector and tensor modes},'' {\em Phys. Rev. D}, vol.~69, p.~063002, 2004.

\bibitem{Ananda:2006af}
K.~N. Ananda, C.~Clarkson, and D.~Wands, ``{The Cosmological gravitational wave
  background from primordial density perturbations},'' {\em Phys. Rev. D},
  vol.~75, p.~123518, 2007.

\bibitem{Baumann:2007zm}
D.~Baumann, P.~J. Steinhardt, K.~Takahashi, and K.~Ichiki, ``{Gravitational
  Wave Spectrum Induced by Primordial Scalar Perturbations},'' {\em Phys. Rev.
  D}, vol.~76, p.~084019, 2007.

\bibitem{Domenech:2021ztg}
G.~Dom\`enech, ``{Scalar Induced Gravitational Waves Review},'' {\em Universe},
  vol.~7, no.~11, p.~398, 2021.

\bibitem{Perna:2024ehx}
G.~Perna, C.~Testini, A.~Ricciardone, and S.~Matarrese, ``{Fully non-Gaussian
  Scalar-Induced Gravitational Waves},'' {\em JCAP}, vol.~05, p.~086, 2024.

\bibitem{Iovino:2024sgs}
A.~J. Iovino, S.~Matarrese, G.~Perna, A.~Ricciardone, and A.~Riotto, ``{How
  Well Do We Know the Scalar-Induced Gravitational Waves?},'' 12 2024.

\bibitem{Witten:1984rs}
E.~Witten, ``{Cosmic Separation of Phases},'' {\em Phys. Rev. D}, vol.~30,
  pp.~272--285, 1984.

\bibitem{Hogan:1986qda}
C.~J. Hogan, ``{Gravitational radiation from cosmological phase transitions},''
  {\em Mon. Not. Roy. Astron. Soc.}, vol.~218, pp.~629--636, 1986.

\bibitem{Caprini:2009fx}
C.~Caprini, R.~Durrer, T.~Konstandin, and G.~Servant, ``{General Properties of
  the Gravitational Wave Spectrum from Phase Transitions},'' {\em Phys. Rev.
  D}, vol.~79, p.~083519, 2009.

\bibitem{Kamionkowski:1993fg}
M.~Kamionkowski, A.~Kosowsky, and M.~S. Turner, ``{Gravitational radiation from
  first order phase transitions},'' {\em Phys. Rev. D}, vol.~49,
  pp.~2837--2851, 1994.

\bibitem{Caprini:2007xq}
C.~Caprini, R.~Durrer, and G.~Servant, ``{Gravitational wave generation from
  bubble collisions in first-order phase transitions: An analytic approach},''
  {\em Phys. Rev. D}, vol.~77, p.~124015, 2008.

\bibitem{Huber:2008hg}
S.~J. Huber and T.~Konstandin, ``{Gravitational Wave Production by Collisions:
  More Bubbles},'' {\em JCAP}, vol.~09, p.~022, 2008.

\bibitem{Caprini:2009yp}
C.~Caprini, R.~Durrer, and G.~Servant, ``{The stochastic gravitational wave
  background from turbulence and magnetic fields generated by a first-order
  phase transition},'' {\em JCAP}, vol.~12, p.~024, 2009.

\bibitem{Hindmarsh:2013xza}
M.~Hindmarsh, S.~J. Huber, K.~Rummukainen, and D.~J. Weir, ``{Gravitational
  waves from the sound of a first order phase transition},'' {\em Phys. Rev.
  Lett.}, vol.~112, p.~041301, 2014.

\bibitem{Caprini:2024ofd}
C.~Caprini, O.~Pujol\`as, H.~Quelquejay-Leclere, F.~Rompineve, and D.~A. Steer,
  ``{Primordial gravitational wave backgrounds from phase transitions with next
  generation ground based detectors},'' 6 2024.

\bibitem{Nielsen:1973cs}
H.~B. Nielsen and P.~Olesen, ``{Vortex Line Models for Dual Strings},'' {\em
  Nucl. Phys. B}, vol.~61, pp.~45--61, 1973.

\bibitem{Sakellariadou:2009ev}
M.~Sakellariadou, ``{Cosmic Strings and Cosmic Superstrings},'' {\em Nucl.
  Phys. B Proc. Suppl.}, vol.~192-193, pp.~68--90, 2009.

\bibitem{Vilenkin:1981bx}
A.~Vilenkin, ``{Gravitational radiation from cosmic strings},'' {\em Phys.
  Lett. B}, vol.~107, pp.~47--50, 1981.

\bibitem{Sakellariadou:1990ne}
M.~Sakellariadou, ``{Gravitational waves emitted from infinite strings},'' {\em
  Phys. Rev. D}, vol.~42, pp.~354--360, 1990.
\newblock [Erratum: Phys.Rev.D 43, 4150 (1991)].

\bibitem{Hindmarsh:1990xi}
M.~Hindmarsh, ``{Gravitational radiation from kinky infinite strings},'' {\em
  Phys. Lett. B}, vol.~251, pp.~28--33, 1990.

\bibitem{Damour:2004kw}
T.~Damour and A.~Vilenkin, ``{Gravitational radiation from cosmic
  (super)strings: Bursts, stochastic background, and observational windows},''
  {\em Phys. Rev. D}, vol.~71, p.~063510, 2005.

\bibitem{Thrane:2013oya}
E.~Thrane and J.~D. Romano, ``{Sensitivity curves for searches for
  gravitational-wave backgrounds},'' {\em Phys. Rev. D}, vol.~88, no.~12,
  p.~124032, 2013.

\bibitem{saxony_einstein_telescope_2024}
E.~T. EMR, ``Saxony also wants to build the einstein telescope,'' December
  2024.

\bibitem{Iacovelli:2024mjy}
F.~Iacovelli, E.~Belgacem, M.~Maggiore, M.~Mancarella, and N.~Muttoni,
  ``{Combining underground and on-surface third-generation gravitational-wave
  interferometers},'' {\em JCAP}, vol.~10, p.~085, 2024.

\bibitem{Ebersold:2024hgp}
M.~Ebersold, T.~Regimbau, and N.~Christensen, ``{Next-generation global
  gravitational-wave detector network: Impact of detector orientation on
  compact binary coalescence and stochastic gravitational-wave background
  searches},'' 8 2024.

\bibitem{Koley:2022wpe}
S.~Koley, M.~Bader, J.~van~den Brand, X.~Campman, H.~J. Bulten, F.~Linde, and
  B.~Vink, ``{Surface and underground seismic characterization at Terziet in
  Limburg\textemdash{}the Euregio Meuse\textendash{}Rhine candidate site for
  Einstein Telescope},'' {\em Class. Quant. Grav.}, vol.~39, no.~2, p.~025008,
  2022.

\bibitem{Saulson:1984yg}
P.~R. Saulson, ``{Terrestrial Gravitational Noise On A Gravitational Wave
  Antenna},'' {\em Phys. Rev. D}, vol.~30, pp.~732--736, 1984.

\bibitem{Badaracco:2019vjq}
F.~Badaracco and J.~Harms, ``{Optimization of seismometer arrays for the
  cancellation of Newtonian noise from seismic body waves},'' {\em Class.
  Quant. Grav.}, vol.~36, no.~14, p.~145006, 2019.

\bibitem{Janssens:2021cta}
K.~Janssens, K.~Martinovic, N.~Christensen, P.~M. Meyers, and M.~Sakellariadou,
  ``{Impact of Schumann resonances on the Einstein Telescope and projections
  for the magnetic coupling function},'' {\em Phys. Rev. D}, vol.~104, no.~12,
  p.~122006, 2021.
\newblock [Erratum: Phys.Rev.D 105, 109904 (2022)].

\bibitem{zhu2017asymptotic}
Z.~Zhu and M.~B. Wakin, ``{On the asymptotic equivalence of circulant and
  Toeplitz matrices},'' {\em IEEE Transactions on Information Theory}, vol.~63,
  no.~5, pp.~2975--2992, 2017.

\bibitem{Hartwig:2023pft}
O.~Hartwig, M.~Lilley, M.~Muratore, and M.~Pieroni, ``{Stochastic gravitational
  wave background reconstruction for a nonequilateral and unequal-noise LISA
  constellation},'' {\em Phys. Rev. D}, vol.~107, no.~12, p.~123531, 2023.

\bibitem{Kume:2024sbu}
J.~Kume, M.~Peloso, M.~Pieroni, and A.~Ricciardone, ``{Assessing the Impact of
  Unequal Noises and Foreground Modeling on SGWB Reconstruction with LISA},''
  10 2024.

\bibitem{Liu:2024jna}
J.~Liu, A.~Vajpeyi, R.~Meyer, K.~Janssens, J.~E. Lee, P.~Maturana-Russel,
  N.~Christensen, and Y.~Liu, ``{Variational inference for correlated
  gravitational wave detector network noise},'' 9 2024.

\bibitem{Amann:2020jgo}
F.~Amann {\em et~al.}, ``{Site-selection criteria for the Einstein
  Telescope},'' {\em Rev. Sci. Instrum.}, vol.~91, no.~9, p.~9, 2020.

\bibitem{Wong:2024hes}
I.~C.~F. Wong, P.~T.~H. Pang, M.~Wils, F.~Cireddu, W.~Del~Pozzo, and T.~G.~F.
  Li, ``{The Potential Impact of Noise Correlation in Next-generation
  Gravitational Wave Detectors},'' 7 2024.

\bibitem{LIGOScientific:2016gtq}
B.~P. Abbott {\em et~al.}, ``{Characterization of transient noise in Advanced
  LIGO relevant to gravitational wave signal GW150914},'' {\em Class. Quant.
  Grav.}, vol.~33, no.~13, p.~134001, 2016.

\bibitem{aLIGO:2020wna}
A.~Buikema {\em et~al.}, ``{Sensitivity and performance of the Advanced LIGO
  detectors in the third observing run},'' {\em Phys. Rev. D}, vol.~102, no.~6,
  p.~062003, 2020.

\bibitem{LIGOScientific:2014qfs}
J.~Aasi {\em et~al.}, ``{Characterization of the LIGO detectors during their
  sixth science run},'' {\em Class. Quant. Grav.}, vol.~32, no.~11, p.~115012,
  2015.

\bibitem{LIGO:2024kkz}
S.~Soni {\em et~al.}, ``{LIGO Detector Characterization in the first half of
  the fourth Observing run},'' 9 2024.

\bibitem{Cornish:2020odn}
N.~J. Cornish, ``{Time-Frequency Analysis of Gravitational Wave Data},'' 8
  2020.

\bibitem{LIGOScientific:2019vic}
B.~P. Abbott {\em et~al.}, ``{Search for the isotropic stochastic background
  using data from Advanced LIGO\textquoteright{}s second observing run},'' {\em
  Phys. Rev. D}, vol.~100, no.~6, p.~061101, 2019.

\bibitem{KAGRA:2021kbb}
R.~Abbott {\em et~al.}, ``{Upper limits on the isotropic gravitational-wave
  background from Advanced LIGO and Advanced Virgo\textquoteright{}s third
  observing run},'' {\em Phys. Rev. D}, vol.~104, no.~2, p.~022004, 2021.

\bibitem{Armstrong_1999}
J.~Armstrong, F.~Estabrook, and M.~Tinto, ``{Time-Delay Interferometry for
  Space-based Gravitational Wave Searches},'' {\em The Astrophysical Journal},
  vol.~527, p.~814, dec 1999.

\bibitem{Ashton:2018jfp}
G.~Ashton {\em et~al.}, ``{BILBY: A user-friendly Bayesian inference library
  for gravitational-wave astronomy},'' {\em Astrophys. J. Suppl.}, vol.~241,
  no.~2, p.~27, 2019.

\bibitem{Romero-Shaw:2020owr}
I.~M. Romero-Shaw {\em et~al.}, ``{Bayesian inference for compact binary
  coalescences with bilby: validation and application to the first
  LIGO\textendash{}Virgo gravitational-wave transient catalogue},'' {\em Mon.
  Not. Roy. Astron. Soc.}, vol.~499, no.~3, pp.~3295--3319, 2020.

\bibitem{Morisaki:2023kuq}
S.~Morisaki, R.~Smith, L.~Tsukada, S.~Sachdev, S.~Stevenson, C.~Talbot, and
  A.~Zimmerman, ``{Rapid localization and inference on compact binary
  coalescences with the Advanced LIGO-Virgo-KAGRA gravitational-wave detector
  network},'' {\em Phys. Rev. D}, vol.~108, no.~12, p.~123040, 2023.

\bibitem{Speagle:2019ivv}
J.~S. Speagle, ``{dynesty: a dynamic nested sampling package for estimating
  Bayesian posteriors and evidences},'' {\em Mon. Not. Roy. Astron. Soc.},
  vol.~493, no.~3, pp.~3132--3158, 2020.

\bibitem{Capurri:2021zli}
G.~Capurri, A.~Lapi, C.~Baccigalupi, L.~Boco, G.~Scelfo, and T.~Ronconi,
  ``{Intensity and anisotropies of the stochastic gravitational wave background
  from merging compact binaries in galaxies},'' {\em JCAP}, vol.~11, p.~032,
  2021.

\bibitem{Bellomo:2021mer}
N.~Bellomo, D.~Bertacca, A.~C. Jenkins, S.~Matarrese, A.~Raccanelli,
  T.~Regimbau, A.~Ricciardone, and M.~Sakellariadou, ``{CLASS\_GWB: robust
  modeling of the astrophysical gravitational wave background anisotropies},''
  {\em JCAP}, vol.~06, no.~06, p.~030, 2022.

\bibitem{Boco:2019teq}
L.~Boco, A.~Lapi, S.~Goswami, F.~Perrotta, C.~Baccigalupi, and L.~Danese,
  ``{Merging Rates of Compact Binaries in Galaxies: Perspectives for
  Gravitational Wave Detections},'' 7 2019.

\bibitem{Perigois:2021ovr}
C.~P\'erigois, F.~Santoliquido, Y.~Bouffanais, U.~N. Di~Carlo, N.~Giacobbo,
  S.~Rastello, M.~Mapelli, and T.~Regimbau, ``{Gravitational background from
  dynamical binaries and detectability with 2G detectors},'' {\em Phys. Rev.
  D}, vol.~105, no.~10, p.~103032, 2022.

\bibitem{Dvorkin:2016wac}
I.~Dvorkin, E.~Vangioni, J.~Silk, J.-P. Uzan, and K.~A. Olive,
  ``{Metallicity-constrained merger rates of binary black holes and the
  stochastic gravitational wave background},'' {\em Mon. Not. Roy. Astron.
  Soc.}, vol.~461, no.~4, pp.~3877--3885, 2016.

\bibitem{Cusin:2019jpv}
G.~Cusin, I.~Dvorkin, C.~Pitrou, and J.-P. Uzan, ``{Properties of the
  stochastic astrophysical gravitational wave background: astrophysical sources
  dependencies},'' {\em Phys. Rev. D}, vol.~100, no.~6, p.~063004, 2019.

\bibitem{Planck:2018vyg}
N.~Aghanim {\em et~al.}, ``{Planck 2018 results. VI. Cosmological
  parameters},'' {\em Astron. Astrophys.}, vol.~641, p.~A6, 2020.
\newblock [Erratum: Astron.Astrophys. 652, C4 (2021)].

\bibitem{Martini:2024daa}
A.~Martini, S.~Schmidt, G.~Ashton, and W.~Del~Pozzo, ``{Maximum entropy
  spectral analysis: an application to gravitational waves data analysis},''
  {\em Eur. Phys. J. C}, vol.~84, no.~10, p.~1023, 2024.

\bibitem{GWBird}
I.~Caporali and A.~Ricciardone, ``\texttt{GWBird}: A tool for the detection of
  the stochastic gravitational wave background for current and next generation
  interferometers.,'' {\em To appear}.

\bibitem{Christensen:2018iqi}
N.~Christensen, ``{Stochastic Gravitational Wave Backgrounds},'' {\em Rept.
  Prog. Phys.}, vol.~82, no.~1, p.~016903, 2019.

\bibitem{1626a8bf-8479-36bf-b395-22e4def81091}
J.~Wishart, ``The generalised product moment distribution in samples from a
  normal multivariate population,'' {\em Biometrika}, vol.~20A, no.~1/2,
  pp.~32--52, 1928.

\bibitem{Cameron:2010bh}
E.~Cameron, ``{On the Estimation of Confidence Intervals for Binomial
  Population Proportions in Astronomy: The Simplicity and Superiority of the
  Bayesian Approach},'' {\em Publ. Astron. Soc. Austral.}, vol.~28, p.~128,
  2011.

\end{thebibliography}

\end{document}